\newcommand{\arxiv}[1]{#1} %
\newcommand{\lipics}[1]{} %
\newcommand{\popl}[1]{} %
\newcommand{\lncs}[1]{} %
\newcommand{\icdcs}[1]{} %
\newcommand{\noticdcs}[1]{#1} %
\newcommand{\ppdp}[1]{} %
\newcommand{\notppdp}[1]{#1} %

\newcommand{\onecol}[1]{#1}
\newcommand{\twocol}[1]{}

\newcommand{\anonymous}[1]{}
\newcommand{\notanonymous}[1]{#1}

\newcommand{\Hex}[1]{\hspace{#1ex}}
\newcommand{\Vex}[1]{\vspace{#1ex}}

\ppdp{}

\arxiv{
\documentclass[11pt]{article}
\usepackage{fullpage}
}

\icdcs{}

\usepackage{url}
\usepackage{xspace} 
\usepackage{alltt}
\usepackage[T1]{fontenc}
\usepackage{ifpdf}
\usepackage{graphicx}

\usepackage{listings}
\usepackage{setspace} %
\usepackage{anyfontsize}
\usepackage{t1enc} %
\usepackage{xcolor}
\definecolor{codegreen}{rgb}{0,0.6,0}
\definecolor{codegray}{rgb}{0.5,0.5,0.5}
\definecolor{codepurple}{rgb}{0.58,0,0.82}
\lstdefinestyle{mystyle}{
    backgroundcolor=\color{white},   
    basicstyle=\scriptsize\ttfamily,
    commentstyle=\color{codegreen},
    keywordstyle=\color{magenta},
    stringstyle=\color{codepurple},
    numberstyle=\tiny\color{codegray},
    breakatwhitespace=false,
    breaklines=false,                 
    captionpos=b,
    keepspaces=true,                 
    numbers=none,
    numbersep=5pt,                  
    showspaces=false,                
    showstringspaces=false,
    showtabs=false,                  
    tabsize=2
}
\newenvironment{code}{\begin{alltt}\fontsize{8}{8}\selectfont}{\end{alltt}}
\notppdp{\newcommand\co[1]{\mbox{\tt\small #1}}} %

\ppdp{} %
\newcommand\p[1]{\m{#1}}
\newcommand\m[1]{\mbox{$#1$}} %
\newcommand{\defn}[1]{\textit{#1}} %
\def\mathify#1{\ifmmode{\mbox{$#1$}}\else\mbox{$#1$}\fi}

\newcommand{\mysec}[1]{\section{#1}}
\arxiv{\newcommand{\mypar}[1]{\subsection{#1}}}
\icdcs{}
\ppdp{}

\arxiv{\newcommand{\mysubpar}[1]{\Vex{-.5}\paragraph*{#1.}}}
\lipics{}
\popl{}
\lncs{}
\icdcs{}
\ppdp{}

\usepackage{enumitem}
\arxiv{\setlist[description]{labelindent=2ex}}
\lipics{}
\popl{}
\lncs{}
\lncs{}
\lncs{}
\icdcs{}
\icdcs{}
\icdcs{}
\ppdp{}

\usepackage{boxedminipage}

\usepackage{tikz} 
\usetikzlibrary{arrows,positioning,decorations.pathreplacing,decorations.markings} 
\tikzset{
    >=stealth',
    varbox/.style={
           rectangle, rounded corners, draw=black, thick,
           text width=2.3em, minimum height=1.5em,
           text centered},
    clause/.style={
           <-, thin,
           shorten <=2pt,
           shorten >=2pt},
}
\tikzset{->-/.style={decoration={
  markings,
  mark=at position #1 with {\arrow{>}}},postaction={decorate}}
}
\tikzset{-dot-/.style={decoration={
  markings,
  mark=at position #1 with {\arrow{*}}},postaction={decorate}}
}

\newcommand{\notes}[1]{} %
\newcommand{\more}[1]{} %

\begin{document}

\newcommand{\ackgrants}{This work was supported in part by NSF under
    grants CCF-1414078, %
    CCF-1248184, %
    and CNS-1421893 %
    and ONR under grant 
    N000141512208. %
}

\newcommand{\contact}{
Contact Author:
Y. Annie Liu, 
Computer Science Department, 
Stony Brook University,
Stony Brook, NY 11794-2424,
U.S.A.
liu@cs.stonybrook.edu. 
}

\arxiv{
\title{Moderately Complex Paxos Made Simple:\\
  High-Level Executable Specification of Distributed Algorithms%
  \footnote{\ackgrants \contact}
}
\author{Yanhong A. Liu \Hex{8} Saksham Chand \Hex{8}
Scott D. Stoller\bigskip\\
Computer Science Department, Stony Brook University%
}
\date{}
\maketitle
} %

\lipics{} %

\popl{} %

\lncs{} %

\icdcs{} %

\ppdp{} %

\Vex{-2}
\begin{abstract}

This paper describes the application of a high-level language and method in
developing simpler specifications of more complex variants of the Paxos
algorithm for distributed consensus.  The specifications are for
Multi-Paxos with preemption, replicated state machine, and reconfiguration
and optimized with state reduction and failure detection.  The language is
DistAlgo.
The key is to express complex control flows and synchronization conditions
precisely at a high level, using nondeterministic waits and message-history
queries.  We obtain complete executable specifications that are almost
completely declarative---updating only a number for the protocol round
besides the sets of messages sent and received.

We show the following results: (1) English and pseudocode descriptions of
distributed algorithms can be captured %
completely and precisely at a high level,
without adding, removing, or reformulating algorithm details to fit
lower-level, more abstract, or less direct languages.  (2) We created
higher-level control flows and synchronization conditions than all previous
specifications, and obtained specifications that are much simpler and
smaller, even matching or smaller than abstract specifications that omit
many algorithm details.  (3) The simpler specifications led us to easily
discover useless replies, unnecessary delays, and liveness violations (if
messages can be lost) %
in previous published specifications, by just following the simplified
algorithm flows. (4) The resulting specifications can be executed directly,
and we can express optimizations cleanly, yielding drastic performance
improvement over naive execution and facilitating a general method for
merging processes.
(5) We %
systematically translated the resulting specifications into TLA+ and
developed machine-checked safety proofs, %
which also allowed us to detect and fix a subtle safety violation 
in an earlier unpublished specification.
Additionally, we show the basic concepts in Paxos that are fundamental in
many distributed algorithms and show that they are captured concisely in
our specifications.

\notes{}
\end{abstract}

\notes{}

\popl{}
\ppdp{}

\notes{} %

\mysec{Introduction}
\label{sec-intro}

Distributed algorithms are increasingly important as distributed
applications are increasingly needed.  These algorithms describe how
distributed processes compute and communicate with each other by passing
messages. %
However, because processes are concurrent and can fail and messages can be
lost, delayed, reordered, and duplicated, distributed algorithms are often
difficult to understand, even if they might appear simple.
The most important and well-known of such algorithms is
Paxos~\cite{Lam98paxos,lam01paxos,vra15paxos} for \defn{distributed
  consensus}---a set of distributed processes trying to agree on a single
value or a continuing sequence of values, called \defn{single-value
  consensus} or \defn{multi-value consensus}, respectively.

\mysubpar{Distributed consensus and Paxos}
Distributed consensus is essential in any important service that must
maintain a state, %
including many services provided by companies like Google and Amazon.  This
is because such services must use replication to tolerate failures caused
by machine crashes, network outage, etc. %
Replicated processes must agree on the state of the service or the sequence
of operations that have been performed, e.g., a customer order has been
placed and paid but not yet shipped, so that when some processes become
unavailable, the remaining processes can continue to function correctly.
Distributed consensus is exactly this agreement problem.

Many algorithms and variations have been proposed for distributed
consensus, starting from Virtual Synchrony (VS) by Birman and
Joseph~\cite{birman1987reliable} and Viewstamped Replication (VR) by Oki
and Liskov~\cite{oki88vsr}.  These algorithms share similar ideas, but
Paxos became the most well-known name and the focus of studies 
starting with its elaborate description by
Lamport~\cite{Lam98paxos,lam-paxos-history}.
The basic ideas %
in these algorithms are fundamental for not only
consensus, but all distributed algorithms that must deal with replication,
asynchronous communication, and failures.  These ideas include successive
rounds, a.k.a.\ views in VR~\cite{oki88vsr} and ballots in
Paxos~\cite{Lam98paxos}, leader election; voting by majority or quorum;
preemption; dynamic reconfiguration; state reduction; and failure detection
by periodic probe or heartbeat.  Van Renesse et al.\ give extended
discussions of these ideas~\cite{vra15paxos} and comparisons of major
variants of Paxos~\cite{van15vive}. %

Paxos is well known to be hard to
understand~\cite{lam01paxos}. %
Since Paxos was introduced by Lamport~\cite{Lam98paxos,lam-paxos-history},
there has been a continuous series of studies of it.
This includes not only optimizations and extensions, especially for use in
practical systems, e.g., Google's distributed lock service
Chubby~\cite{burrows06chubby,chandra07paxos},
but also more variations and expositions, especially with effort for better
understanding,
e.g.,~\cite{lampson1996build,prisco00revisit,lam01paxos,ongaro14raft,vra15paxos,van15vive,garcia18paxos},
and for formal verification,
e.g.,~\cite{hawblitzel2015ironfleet,wilcox2015verdi,woos2016planning,Cha+16PaxosTLAPS-FM,padon2017paxos,ChaLiu18PaxosHistVarTLAPS-NFM,taube2018modular}.
Major developments in this series have led to a better comprehensive
understanding of Paxos, as presented in Paxos Made Moderately Complex by
van Renesse and Altinbuken~\cite{vra15paxos}, starting from its simpler
core, as presented in Paxos Made Simple by Lamport~\cite{lam01paxos}.
Can these algorithms be made completely precise, readily executable in real
distributed environments, and at the same time easier to understand?

\mysubpar{This paper}
This paper describes the application of a high-level language and method in
developing simpler specifications of both basic and more complex variants
of the Paxos algorithm for distributed consensus.
The specifications are for Paxos for single-value consensus, as described
by Lamport in English~\cite{lam01paxos}, which we call \defn{Basic Paxos},
and Paxos for multi-value consensus with preemption, replicated state
machine, and reconfiguration, as described by van Renesse and Altinbuken in
pseudocode~\cite{vra15paxos}, which we call \defn{vRA Multi-Paxos}, as well
as vRA Multi-Paxos optimized with state reduction and failure
detection~\cite{vra15paxos}.  The key is to express complex control flows
and synchronization conditions precisely at a high level, using
nondeterministic waits and message-history queries.  We obtain complete
executable specifications that are almost completely declarative---updating
only a number for the protocol round besides the sets of messages sent and
received.
Our contributions include the following:
\lipics{}
\begin{itemize}
  \arxiv{\setlength{\itemsep}{0ex}}

\item We show that English and pseudocode descriptions of algorithms
  can be captured completely and precisely at a high level,
  without adding, removing, or reformulating algorithm details to fit
  lower-level, more abstract, or less direct languages.

\item We created higher-level control flows and synchronization conditions
  than all previous specifications of these algorithms, and obtained
  specifications that are much simpler and smaller,
  even matching or smaller than abstract specifications that omit many
  algorithm details.

\item We show that the simpler specifications led us to easily discover
  useless replies, unnecessary delays, and liveness violations (if messages
  can be lost) in the original vRA Multi-Paxos specification and in our
  earlier specification~\cite{Liu+12DistSpec-SSS}, by just following the
  simplified algorithm flows.

\item We demonstrate that the resulting specifications can be executed
  directly, and we can express optimizations cleanly,
  yielding drastic performance improvement over naive execution
  and facilitating a general method for merging processes.

\notes{}

\item We %
systematically translated the resulting specifications into\ppdp{}
TLA+~\cite{Lam02book} and developed
machine-checked safety proofs, 
which also allowed us to detect and fix a subtle safety violation in an
earlier unpublished specification.
\end{itemize}
We also show the basic concepts in Paxos that are fundamental in many
distributed algorithms and show that they are captured concisely in our
specifications.

Our specifications are written in DistAlgo~\cite{distalgo17lang}, a language
for distributed algorithms with a formal operational
semantics~\cite{Liu+17DistPL-TOPLAS} and a complete implementation in 
Python~\cite{Liu+12DistPL-OOPSLA,distalgo18git}. 
Our complete executable specifications in DistAlgo and machine-checked
proofs in TLAPS are available at
\anonymous{}%
\notanonymous{\url{darlab.cs.stonybrook.edu/paxos}}.

There have been numerous studies of specifications for understanding and
verification of distributed algorithms, especially of Paxos, as discussed
in Section~\ref{sec-related}.  With the exception of vRA Multi-Paxos, no
previous papers present direct, complete, and precise specification of
Multi-Paxos.  Previous formal specifications are only for Basic Paxos,
e.g.,~\cite{kellomaki2004ann,BudSandboxPaxos}, %
are abstract by omitting many algorithm details necessary for real
execution, e.g.,~\cite{Cha+16PaxosTLAPS-FM,padon2017paxos}, or are too long
or too complicated to include in papers,
e.g.,~\cite{hawblitzel2015ironfleet,woos2016planning}.  Also, to the best
of our knowledge, no previous efforts of specification and formal
verification, for any Paxos variant, reported finding any correctness
violations %
or improvements. %
Nor did previous efforts of implementation of vRA Multi-Paxos in various
languages, including in Python~\cite{vra15paxos}.

\mysec{Basic Paxos, language, and high-level specification}
\label{sec-basic}

We describe the language, DistAlgo, and method of specification using Basic
Paxos as an example.  Prior knowledge of Paxos can be helpful, but the
description is self-contained.  

Paxos considers distributed processes that may crash and may later recover,
and messages that may be lost, delayed, reordered, and duplicated.  It
guarantees \defn{safety}, i.e., agreement on the decided single value in
Basic Paxos (or sequence of values in Multi-Paxos) by nonfaulty processes,
and validity of the decided value (or values) to be among allowed values.
However, it does not guarantee \defn{liveness}, i.e., nonfaulty processes
eventually decide, without stronger assumptions, due to the well-known
impossibility result~\cite{fischer85flp}.

\mypar{Basic Paxos in English}
\label{sec-basic-english}

Figure~\ref{fig-lapaxos-paper} shows Lamport's description of Basic Paxos
in English~\cite{lam01paxos}.
It presents the algorithm for (1) the proposer and acceptor---the two
phases\more{}, and
(2) the learner---the obvious algorithm.\more{}
\begin{figure*}[htbp]
  \centering

\fbox{$\!$ 
\fontsize{9}{10}\selectfont 
\begin{tabular}{@{}p{\arxiv{0.85}\popl{}\icdcs{}\ppdp{}\textwidth}@{}}

\mbox{\hspace{2.5ex}} 
Putting the actions of the proposer and acceptor together, we see that
the algorithm operates in the following two phases.

\arxiv{\Vex{-2}}
\begin{description}
  \setlength{\itemsep}{.5ex}
  \setlength{\parskip}{1ex}

\item {\bf Phase 1.} (a) A proposer selects a proposal number \m{n} and sends
  a \m{prepare} request with number \m{n} to a majority of acceptors.

  (b) If an acceptor receives a \m{prepare} request with number \m{n}
  greater than that of any \m{prepare} request to which it has already
  responded, then it responds to the request with a promise not to accept
  any more proposals numbered less than \m{n} and with the highest-numbered
  proposal (if any) that it has accepted.

\item {\bf Phase 2.} (a) If the proposer receives a response to its
  \m{prepare} requests (numbered \m{n}) from a majority of acceptors, then
  it sends an \m{accept} request to each of those acceptors for a proposal
  numbered \m{n} with a value \m{v}, where \m{v} is the value of the
  highest-numbered proposal among the responses, or is any value if the
  responses reported no proposals.

  (b) If an acceptor receives an \m{accept} request for a proposal numbered
  \m{n}, it accepts the proposal unless it has already responded to a
  \m{prepare} request having a number greater than \m{n}.

\end{description}
\arxiv{\Vex{-4.5}}
\icdcs{}
\ppdp{}

\more{} %
\\\hline
\Vex{-.5}
\Hex{2.5}
To learn that a value has been chosen, a learner must find out that a proposal
has been accepted by a majority of acceptors. The obvious algorithm
is to have each acceptor, whenever it accepts a proposal, respond to all
learners, sending them the proposal. %
\more{} %
\end{tabular}
} %
\arxiv{\Vex{-1}}
\icdcs{}
\ppdp{}
\caption{\small
Lamport's description of Basic Paxos in English~\cite{lam01paxos}.\Hex{0}}
\label{fig-lapaxos-paper}
\Vex{4}
\icdcs{}
\begin{minipage}[c]{\textwidth}
\rule{1\columnwidth}{.1mm}\notppdp{\Vex{-1}}
\begin{code}
\ppdp{}   1 process Proposer:
\ppdp{}   2   def setup(acceptors):                                  # take in set of acceptors
\ppdp{}   3     self.majority := acceptors                           # any majority of acceptors; we use all
\ppdp{}   4   def run():
\ppdp{}   5     n := self                                                       # Phase 1a: select proposal num n
\ppdp{}   6     send ('prepare',n) to majority                                  #  send prepare n to majority\Vex{0}
\ppdp{}   7     await count \{a: received ('respond',=n,_) from a\}               # Phase 2a: wait for response to n
\ppdp{}   8           > (count acceptors)/2:                                    #  from a majority of acceptors
\ppdp{}   9       v := any (\{v: received ('respond',=n,(n2,v)),                 #  find val in accepted prop, if any
\ppdp{}  10                     n2 = max \{n3: received ('respond',=n,(n3,_))\} \} #  having max proposal num
\ppdp{}  11                 or \{any 1..100\})                                    #  or any value; use any in 1..100      
\ppdp{}  12       responded := \{a: received ('respond',=n,_) from a\}            #  find responded
\ppdp{}  13       send ('accept',n,v) to responded                              #  send accept for proposal n,v\Vex{0} 
\ppdp{}  14 process Acceptor:
\ppdp{}  15   def setup(learners): pass                              # take in set of learners
\ppdp{}  16   def run(): await false                                 # wait for nothing, only to handle messages\Vex{0}
\ppdp{}  17   receive ('prepare',n) from p:                                     # Phase 1b: receive prepare n
\ppdp{}  18     if each sent ('respond',n2,_) has n > n2:                       #  if n > each responded n2
\ppdp{}  19       max_prop := any \{(n,v): sent ('accepted',n,v),                #  find accepted proposal, if any,
\ppdp{}  20                               n = max \{n: sent ('accepted',n,_)\} \}  #  having max proposal num
\ppdp{}  21       send ('respond',n,max_prop) to p                              #  respond with n,max_prop\Vex{0}
\ppdp{}  22   receive ('accept',n,v):                                           # Phase 2b: receive proposal n,v
\ppdp{}  23     if not some sent ('respond',n2,_) has n2 > n:                   #  if not responded with larger n2
\ppdp{}  24       send ('accepted',n,v) to learners                             #  send accepted proposal n,v\Vex{0} 
\ppdp{}  25 process Learner:
\ppdp{}  26   def setup(acceptors): pass                             # take in set of acceptors
\ppdp{}  27   def run():
\ppdp{}  28     await some received ('accepted',n,v) has                        # wait for some accepted proposal 
\ppdp{}  29                count \{a: received ('accepted',=n,=v) from a\}        # that has been accepted by
\ppdp{}  30                > (count acceptors)/2:                               # a majority of acceptors
\ppdp{}  31       output('chosen',v)                                 # output chosen value v
\ppdp{}  32 def main():
\ppdp{}  33   acceptors := 3 new Acceptor                            # create 3 Acceptor processes
\ppdp{}  34   proposers := 3 new Proposer(acceptors)                 # create 3 Proposer procs, pass in acceptors
\ppdp{}  35   learners := 3 new Learner(acceptors)                   # create 3 Learner procs, pass in acceptors
\ppdp{}  36   acceptors.setup(learners)                              # to acceptors, pass in learners
\ppdp{}  37   (acceptors + proposers + learners).start()             # start acceptors, proposers, learners
\end{code}\Vex{-3}\ppdp{}
\rule{1\columnwidth}{.1mm}
\end{minipage}
\arxiv{\Vex{-1}}
\icdcs{}
\ppdp{}

\caption{\small
    A high-level specification of Basic Paxos in DistAlgo, including
    setting up and running
    3 each of Proposer, Acceptor, and Learner processes and 
    outputting the result.\icdcs{}}
\label{fig-lapaxos-da}
\end{figure*}
From it, we can see that high-level specification of distributed algorithms
needs four main concepts:
\begin{enumerate}
  \setlength{\itemsep}{0ex}

\item[1.] {\bf Distributed processes that can send messages.}  In
  Figure~\ref{fig-lapaxos-paper},$\!$ there are proposer, acceptor, and learner
  processes, \m{prepare} and \m{accept} messages from a proposer to an
  acceptor, response messages back from an acceptor, and messages for
  accepted proposals from an acceptor to a learner.

\item[2.] {\bf Control flows for handling received messages.}  In
  Figure~\ref{fig-lapaxos-paper}, messages can be received by acceptors,
  proposers, and learners asynchronously at any time, but processes must
  synchronize by testing and waiting for different conditions on the
  received messages.  Capturing such complex control flows is essential.

\item[3.] {\bf High-level queries for synchronization conditions.} In
  Figure~\ref{fig-lapaxos-paper}, the conditions checked in Phases 1b, 2a, 2b, and the
  learner before taking actions involve sets of many messages sent and
  received.  Capturing such conditions at a high level is the most
  important key to making control flows much clearer and easier to
  understand.

\item[4.] {\bf Configuration for setting up and running.}  This is often
  implicit in descriptions of distributed algorithms.  In
  Figure~\ref{fig-lapaxos-paper}, each process needs to be set up and get a
  hold of other processes with which it needs to communicate.  In general,
  there may also be special configuration requirements, such as use of
  specific logical clocks.

\end{enumerate}

Figure~\ref{fig-lapaxos-da} shows a complete high-level executable
specification of Basic Paxos in DistAlgo, including setting up and running.
It will be explained in examples for the language in Section~\ref{sec-lang}
and discussed in more detail in Section~\ref{sec-basic-understanding}.  The
overall structure has two main aspects:
\begin{itemize}

\item Figure~\ref{fig-lapaxos-paper} corresponds to the body of \co{run} in
  \co{Proposer} (lines 5-13), the two \co{receive} definitions in
  \co{Acceptor} (lines 17-24), and the \co{await} condition in
  \co{Learner} (lines 28-30), %
  including selecting ``a proposal number'' in Phase 1a to be self (line
  5), as is commonly used, %
  and taking ``any value'' in Phase 2a to be any integer in %
  1..100 as the set of allowed values (line 11), for simplicity.

\item The rest %
  puts all together, plus setting up processes, starting them, and
  outputting the result of the execution.

\end{itemize}
Note that Figure~\ref{fig-lapaxos-paper}, and thus
Figure~\ref{fig-lapaxos-da}, specifies only one round of the two phases for
each process.  Section~\ref{sec-basic-understanding} discusses successive
rounds that help increase liveness.

\mypar{Language and high-level specification}
\label{sec-lang}

We use DistAlgo, a language that supports the four main concepts in
Section~\ref{sec-basic-english} by building on an object-oriented
programming language, with a formal operational
semantics~\cite{Liu+17DistPL-TOPLAS}.

Besides the language constructs explained below, commonly used notations in
high-level languages are used for no operation (\co{pass}), assignments
(\co{\p{v}\,:=\,\p{e}}), etc.
Indentation is used for scoping, ``\co{:}'' for separation, and ``\co{\#}''
for comments.

\mysubpar{Distributed processes that can send messages}
A type \co{\p{P}} of distributed processes is defined by \co{process
  \p{P}:~\p{body}}, e.g., lines 1-13 in Figure~\ref{fig-lapaxos-da}.
The body may contain
\begin{itemize}
  \setlength{\itemsep}{0ex}

\item a \co{setup} definition for taking in and setting up the
  values used by a type \co{\p{P}} process, e.g., lines 2-3,

\item a \co{run} definition for running the main control flow of the
  process, e.g., lines 4-13, and

\item \co{receive} definitions for handling received messages, e.g., lines
  17-21. %

\end{itemize}
A process can refer to itself as \co{self}. Expression \co{self.\p{attr}}
(or \co{\p{attr}} when there is no ambiguity) refers to the value of
\co{\p{attr}} in the process.
\co{\p{ps} := \p{n} new \p{P}(\p{args})} creates \co{\p{n}} new processes of
type \co{\p{P}}, optionally passing in the values of \co{\p{args}} to \co{setup},
and assigns the new processes to \co{\p{ps}}, e.g., lines 33 and
34.  %
\co{\p{ps}.setup(\p{args})} sets up processes \co{\p{ps}} using values of
\co{\p{args}}, e.g., line 36, and \co{\p{ps}.start()} starts \co{run()} of
processes \co{\p{ps}}, e.g., line 37.

Processes can send messages: \co{send \p{m} to \p{ps}} sends message
\co{\p{m}} to processes \co{\p{ps}}, e.g., line 6.$\!$ %

\mysubpar{Control flow for handling received messages}
Received messages can be handled both asynchronously, using 
\co{receive} definitions, and synchronously, using \co{await}
statements.
\begin{itemize}
  \setlength{\itemsep}{0ex}

\item A %
  definition, \co{receive \p{m} from \p{p}} handles, at yield points,
  un-handled messages that match \co{\p{m}} from \co{\p{p}}, e.g., lines
  17-21.
  There is a yield point before each \co{await} statement, e.g., line~7,
  for handling messages while waiting.
 The \co{from} clause is optional, e.g., line 22.

\item A statement, \co{await \p{cond_1}:\,\p{stmt_1} or\,...\,or
    \p{cond_k}:\,\p{stmt_k}}\ppdp{} \co{timeout\,\p{t}:\,\p{stmt}}, waits for one of
  \co{\p{cond_1}}, ..., \co{\p{cond_k}} to be true or a timeout after
  period \co{\p{t}}, and then nondeterministically selects one of
  \co{\p{stmt_1}}, ..., \co{\p{stmt_k}}, \co{\p{stmt}} whose conditions are
  true to execute, e.g., lines 7-13.
  Each branch is optional.
  So is the statement in \co{await} with a single branch.
\end{itemize}\arxiv{\Vex{-1}}

\mysubpar{High-level queries for synchronization conditions}
High-level queries can be used over message histories, and patterns can be
used for matching messages.
\begin{itemize}
  \setlength{\parskip}{.5ex}

\item Histories of messages sent and received by a process are kept in
  variables \co{sent} and \co{received}, respectively.
  \co{sent} is updated at each \co{send} statement, by adding each message
  sent.
  \co{received} is updated at yield points, 
  by adding un-handled messages before executing all matching
  \co{receive} definitions.

  Expression \co{sent \m{m} to \p{p}} is equivalent to \co{\p{m} to \p{p} in
    sent}.  It returns true iff a message that matches \co{\p{m} to \p{p}}
  is in \co{sent}.
  The \co{to} clause is optional.
  \co{received \m{m} from \p{p}} is similar.

\item A pattern can be used to match a message, in \co{sent} and
  \co{received}, and by a \co{receive} definition.  A constant value, such as
  \co{'respond'}, or a previously bound variable, indicated with prefix
  \co{=}, in the pattern must match the corresponding components of the
  message.  An underscore matches anything.  Previously unbound variables
  in the pattern are bound to the corresponding components in the matched
  message.

  For example, \co{received ('respond',=n,\_) from a} on line 7 matches
  every message triple in \co{received} whose first two components are
  \co{'respond'} and the value of \co{n}, and binds \co{a} to the sender.

\end{itemize}
A query can be a comprehension, aggregation, or %
quantification over sets or sequences.
\lipics{}
\begin{itemize}
  \setlength{\itemsep}{0ex}

\item A comprehension, \co{\{\p{e}:~\p{v\sb{1}} in \p{s\sb{1}}, ...,
    \p{v\sb{k}} in \p{s\sb{k}}, \p{cond}\}}, where \co{\p{v_i}} can be a
  pattern, returns the set of values of \co{\p{e}} for all combinations of
  values of variables that satisfy all \co{\p{v_i} in \p{s\sb{i}}} clauses
  and condition \co{\p{cond}}, e.g., the comprehension on line~7.

\item An aggregation, \co{\p{agg} \p{s}}, where \co{\p{agg}} is an
  aggregation operator such as \co{count} or \co{max}, returns the value of
  applying \co{\p{agg}} to the set value of \co{s}, e.g., the \co{count}
  query on line 7.

\item An existential quantification,\icdcs{} 
  \co{some \p{v\sb{1}} in \p{s\sb{1}},
    ..., \p{v\sb{k}} in \p{s\sb{k}}}\ppdp{} \co{has \p{cond}}, returns true iff for
  some combinations of values of variables that satisfy all \co{\p{v\sb{i}}
    in \p{s\sb{i}}} clauses, \co{\p{cond}} holds, e.g., the \co{some} query
  on line 23.
  When the query returns true, all variables in the query are bound to
  a combination of satisfying values, called a witness, e.g., \co{n} and
  \co{v} on lines 28-30.

\item A universal quantification,\icdcs{} 
  \co{each \p{v\sb{1}} in \p{s\sb{1}}, ...,
    \p{v\sb{k}} in \p{s\sb{k}}}\ppdp{} \co{has \p{cond}}, returns true iff for all
  combinations of values of variables that satisfy all \co{\p{v\sb{i}} in
    \p{s\sb{i}}} clauses, \co{\p{cond}} holds, e.g., the \co{each} query
  on line 18.

\end{itemize}
Other operations on sets can also be used, in particular:\Vex{-0}
\begin{itemize}
  \setlength{\itemsep}{0ex}

\item \co{any \p{s}} returns any element of set \co{\p{s}} if \co{\p{s}} is
  not empty, and a special value \co{undefined} otherwise.

\item
\co{\p{n_1}..\p{n_2}} returns the set of integers ranging from \co{\p{n_1}}
to \co{\p{n_2}} for \co{\p{n_1\leq n_2}}.

\item
\co{\p{s_1} + \p{s_2}} returns the union of sets \co{\p{s_1}} and \co{\p{s_2}}.
\arxiv{
\,}\lipics{}
\item
\co{\p{s_1} or \p{s_2}} returns \co{\p{s_1}} if \co{\p{s_1}} is not empty, and
\co{\p{s_2}} otherwise.
\end{itemize}

\mysubpar{Configuration for setting up and running}
Configuration for requirements such as %
logical clocks can be specified in a \co{main} definition, e.g., lines
32-37.
Basic Paxos does not have special configuration requirements, besides
setting up and running the processes by calling \co{new}, \co{setup}, and
\co{start} as already described.
In general, \co{new} can have an additional clause \co{at \p{node}}
specifying remote nodes where the created processes will run; the default
is the local node.

\mysubpar{High-level specification of distributed algorithms via
  declarative queries}
The core of DistAlgo supports, besides distributed processes that can send
and receive messages, prominently high-level constructs for expressing
complex control flows and synchronization conditions, using
nondeterministic waits with message-history queries.
These constructs are not supported in other languages for concurrent and
distributed processes, including Erlang~\cite{larson09erlang,erlang} and
languages like CSP~\cite{hoare78csp} and CCS~\cite{milner80ccs}.

Examine the Basic Paxos algorithm specification in
Figure~\ref{fig-lapaxos-da}, in processes for Proposer, Acceptor, and
Learner.  One can see that all variables in assignment statements are
either assigned only once (\co{majority} and \co{n}) or assigned
temporarily to be consumed immediately (\co{v}, \co{responded}, and
\co{max\_prop}).  In other words, these variables are like those bound in
\co{let} expressions in functional languages.  Therefore, there are no
intrinsic state updates besides sending and receiving messages that update
the built-in variables \co{sent} and \co{received}.

Sending and receiving messages are essential in distributed algorithms.
However, the rest of the algorithm can be specified declaratively, by using
high-level queries over \co{sent} and \co{received}, and using them in
\co{await} conditions for synchronization and in \co{if} conditions for
safeguarding.  This is the key to making specifications of distributed
algorithms high-level and declarative, and therefore easier to understand.

\mypar{Understanding Basic Paxos and fundamentals of distributed algorithms}
\label{sec-basic-understanding}

For Basic Paxos, 
we now see how Phases 1 and 2 in Figure~\ref{fig-lapaxos-da} precisely
follow Lamport's description in Figure~\ref{fig-lapaxos-paper}.
\lipics{}
\begin{description}
  \setlength{\itemsep}{0ex}

\item {\bf Phase 1a} (lines 5-6). %
  This straightforwardly follows the description in
  Figure~\ref{fig-lapaxos-paper}.  
  \noticdcs{A proposal number can be any value that
    allows the comparison operation $>$.}

\item {\bf Phase 1b} (lines 17-21).  When an acceptor receives a
  \co{prepare} message with a proposal number larger than all numbers
  in its previous responses, it responds with this proposal number and
  with any \co{(n,v)} pair in sent \co{accepted} messages where \co{n}
  is maximum among all such pairs; note that, if it has not sent any
  \co{accepted} messages, \co{max\_prop} is \co{undefined}, instead of
  some \co{(n,v)}, in the \co{respond} message.

\item {\bf Phase 2a} (lines 7-13).  When a proposer receives responses
  to %
  its proposal number \co{n} from a majority of acceptors, it takes \co{v}
  in the \co{(n2,v)} that has the maximum \co{n2} in all responses to
  \co{n}, or any value in allowed values 1..100 if the responses contain no
  \co{(n2,v)} pairs but only \co{undefined}; note that, in the latter case,
  the set on lines 9-10 is empty because \co{undefined} does not match
  \co{(n2,v)}.  The proposer then sends \co{accept} for a proposal with
  number \co{n} and value \co{v} to acceptors that responded to \co{n}.

\item {\bf Phase 2b} (lines 22-24).  This directly follows the
  description in Figure~\ref{fig-lapaxos-paper}.

\end{description}\lipics{}
In particular, the specification in Figure~\ref{fig-lapaxos-da} makes the
``promise'' in Phase 1b of Figure~\ref{fig-lapaxos-paper} precise---the
``promise'' refers to the responded proposal number that will be used later
to not accept any proposal with a smaller proposal number in Phase 2b.

Indeed, this is the hardest part for understanding Paxos, because
understanding the \co{respond} message sent in Phase 1b requires
understanding the \co{accepted} message sent in Phase 2b, a later
phase, but understanding the later phase depends on understanding the
earlier phase.
The key idea is that the Phase 2b to be understood is for a smaller%
\notes{} 
proposal number, i.e., the \co{accepted} messages used in Phase 1b are
those sent with smaller proposal numbers than the \co{n} in received
\co{('prepare', n)}.  For the smallest \co{n}, no \co{accepted} messages
have been sent, and Phase 1b responds with \co{max\_prop} being
\co{undefined}.

Overall, agreement is ensured because (1) a value \co{v} is chosen
only if there is an \co{n} such that the pair (\co{n}, \co{v}) is
accepted by a majority of acceptors (lines 28-31), and (2) once
(\co{n}, \co{v}) is accepted by a majority of acceptors, this \co{v}
will be in the \co{accept} message for each next greater proposal
number (through lines 19-20 and 9-10) that has received responses from
a majority of acceptors (because two majorities overlap).  Validity is
ensured because any \co{v} in \co{accept} and \co{accepted} messages
is one of the allowed values (1..100 as we use) either directly (from
lines 11 and 22) or indirectly (through \co{respond} messages).

\mysubpar{Fundamental concepts in distributed algorithms}
Basic Paxos contains several concepts that are fundamental in
distributed algorithms and are commonly used:
\begin{itemize}
  \setlength{\itemsep}{0ex}

\item {\bf Leader election}. This is as done in Phase 1 but on only lines
  5-6, 17-18, and 21 (ignoring lines 19-20).
  It is for electing at most one proposer (with its proposal number \co{n})
  at a time as the leader.  It ends with the \co{await} condition on lines
  7-8 taking a majority.  In Basic Paxos, only after receiving responses
  from a majority does a proposer carry out Phase 2 and propose to accept a
  value \co{v}.

\item {\bf Majority or quorum voting}.  This is as done on lines 7-8 and
  29-30 that test with a majority.  It ensures that at most one value is in
  the voting result.  In Basic Paxos, the two votings are for electing a
  leader \co{n} on line 7 and choosing an accepted proposal \co{n,v} on
  line 29.

\item {\bf Successive rounds}.  This is as partially done on lines 5, 18,
  and 23, with larger proposal numbers taking over smaller ones.  Rounds
  are used to make progress with increasingly larger numbers.  To do this
  fully to increase liveness, each proposer may iterate (i.e., repeat the
  body of \co{run} if its \co{await} may fail for any reason), with a
  larger number \co{n} in each iteration.  We will see this in Multi-Paxos
  in Section~\ref{sec-multi}, by using a pair for the proposal number,
  called ballot number there; it is the only intrinsic state variable,
  besides \co{sent} and \co{received}, in Multi-Paxos.

\item {\bf Timeout and failure detection}.  To fundamentally increase
  liveness, each proposer may add a timeout to its \co{await}, or an
  alternative branch in \co{await} to receive messages indicating a
  preemption of the condition originally waited for.  Timeout is the most
  important mechanism to provide liveness in practice. We will see
  preemption in Multi-Paxos in Section~\ref{sec-multi}, and see timeout and
  failure detection in Section~\ref{sec-optimize}.

\item {\bf Selection with maximum or minimum}.  This is as done for
  computing \co{max\_prop} and \co{v} on lines 19-20 and 9-11, by using
  maximum to select over collections.  In Basic Paxos, this is for passing
  on the agreed value (once voted by a majority in a round) to successive
  later rounds.  In general, this can be for making monotonic progress
  or for a very different purpose such as breaking symmetry.

\end{itemize}
These are all made simple and precise by our high-level control flows and
synchronization conditions, especially with declarative queries over
message history variables \co{sent} and \co{received}.

To summarize, uses of high-level control flows and declarative queries
allow our Basic Paxos specification to be at the same high level as
Lamport's English description, while making everything %
precise.
With also precise constructs for setting up and running, the complete
specification is both directly executable in real distributed environments,
by automatic compilation to a programming language,
and ready to be verified, by systematic translation to a verifier language.

\mysec{Multi-Paxos with preemption and reconfiguration, and high-level specification}
\label{sec-multi}

Multi-Paxos extends Basic Paxos to reach consensus on a continuing sequence
of values, instead of a single value.
It is significantly more sophisticated than running Basic Paxos for
each \defn{slot} in the sequence, %
because proposals must be made continually for each next slot, with
the proposal number, also called \defn{ballot number} or simply
ballot, incremented repeatedly in new rounds if needed, and the ballot
is shared for as many slots as possible for obvious efficiency reasons.

\defn{Preemption} allows a proposer, also called leader, to be preempted by
another leader that has a larger ballot, i.e., if a leader receives a
message with a larger ballot than its own ballot, it abandons its own
and uses a larger ballot later.

\defn{Replicated state machine} keeps the state of an application, e.g., a
bank.  It continually sends new values, also called commands, to be agreed
on to the leaders, and applies the decided commands to the state of the
application, in order of the slot decided for each command.

\defn{Reconfiguration} allows switching to a set of new leaders during
execution of the algorithm.  The slot in which the change of leaders is to
happen must be agreed on by the old leaders.  This is done by taking the
change as one of the commands to be agreed on.
Note that the new leaders can be set up with a new set of acceptors.

We present a complete specification of Multi-Paxos with preemption,
replicated state machine, and reconfiguration, developed based on vRA
Multi-Paxos.  The specification improves over the original pseudocode in
several ways and also led us to easily discover liveness violations when
messages can be lost.

\mypar{vRA Multi-Paxos pseudocode}

vRA Multi-Paxos gives complete pseudocode for Multi-Paxos with preemption,
replicated state machine, and reconfiguration~\cite{vra15paxos}.
\notes{}
The core ideas are the same as Basic Paxos.
However, except for the name Acceptor, all other names used, for processes,
data, and message types, are %
changed:\lncs{}\icdcs{}
\begin{center}
\small
\onecol{
\begin{tabular}{@{~}c@{~}|@{~}p{14.5ex}@{~}|@{~}p{9ex}@{~}|@{~}c@{~}|@{~}c@{~}|@{~}c@{~}|@{~}c@{~}|@{~}c@{~}}
}
\icdcs{}
\ppdp{}
  Algo\-rithm 
  & Process Types
  & \multicolumn{2}{c}{Data Types~~~}
  & \multicolumn{4}{@{\Hex{\notppdp{-.825}\ppdp{}}}|c}{Message Types}\\\hline
  Basic Paxos
  &Proposer, \mbox{Learner}
  &proposal number
\onecol{
  &value    &\co{prepare}  &\co{respond}  &\co{accept}  &\co{accepted}
}
\icdcs{}
\ppdp{}
  \\\hline
  vRA Multi-Paxos
  &Leader, \mbox{Scout}, Commander 
  &ballot number    
  &com\-mand  &\co{1a}       &\co{1b}       &\co{2a}      &\co{2b}\\\hline
\end{tabular}
\end{center}\ppdp{}
or new:\icdcs{}\ppdp{}
\begin{center}
\small
\onecol{
\begin{tabular}{@{~}c@{~}|@{~}c@{~}|@{~}c@{~}|@{~}c@{~}|@{~}c@{~}|@{~}c@{~}|@{~}c@{~}|@{~}c@{~}}}
\icdcs{}
\ppdp{}
  Algo\-rithm 
  & Process Types & Data Types & \multicolumn{5}{c}{Message Types}\\\hline
  vRA Multi-Paxos
\onecol{%
  & Replica & slot & \co{request} & \co{response} & \co{propose} & 
  \co{decision} & \co{preempt}
}
\icdcs{}
\ppdp{}
\\\hline
\end{tabular}
\end{center}
\begin{itemize}
  \arxiv{\setlength{\itemsep}{-.25ex}}

\item 
An Acceptor process should have the same role as in Basic Paxos,
except that it needs to handle slots.  However, it
is also changed to always reply to \co{1a} and \co{1b} messages without
checking the corresponding conditions as in Basic Paxos; similar conditions 
are checked in Scout and Commander processes to report preemption.

\item
A Leader process spawns Scout and Commander processes to perform Phases 1
and 2, respectively, on its behalf; the spawned processes also determine
preemption and inform the Leader process (\co{preempt} message).  These three
kinds of processes together have essentially the same role as Proposer and
Learner, except that they also handle slots.

\item
A Replica process %
realizes replicated state machine and keeps the state of an application.
It repeatedly receives a
requested command (\co{request}) from a client, and sends a proposed slot
for the command (\co{propose}) to Leader processes; it also receives a
decided slot for a command
(\co{decision}), %
applies the operation in the command to the state at the decided slot, and
sends the result (\co{response}) to the client.

\item
A slot is just a component in a proposal or decision in Leader and
Acceptor, but is tracked in Replica using variables \co{slot\_in} and
\co{slot\_out}, for the next slot to send a proposal and the next slot to
apply a decision, respectively.  
A window between \co{slot\_in} and \co{slot\_out} is used so that a decided
reconfiguration at a slot %
takes effect at the slot %
at the end of the window, while other commands can still be proposed and
decided for slots within the window.

\item
A command is a triple of client id, command id, and operation, and the
operation for reconfiguration holds the set of new leaders.

\end{itemize}
The complete pseudocode (Figs.\,1, 4, 6, 7, and the two formulas
on pages 6, 9, 12, 13, and 14 in~\cite{vra15paxos}) is precise and
succinct, even though not directly executable.  
Appendix~\ref{app-vrapaxos} shows the pseudocode for the Leader process,
the center of the algorithm.
However, there are two main challenges in understanding the overall algorithm:
\begin{enumerate}
  \setlength{\itemsep}{0ex}

\item[1.] Each of the 5 kinds of processes maintains additional process state
  variables.  These variables are updated repeatedly or in multiple places
  in the process without explicit invariants or properties about the values
  in the variables.

\item[2.] Each of the 5 kinds of processes contains an infinite loop.  The body
  of the loop is driven by receiving a message and performing the
  associated actions without expressing higher-level conditions over the
  messages sent or received.

\end{enumerate}
For example, a Replica process maintains 3 sets: \m{\it requests}, \m{\it
  proposals}, and \m{decisions}.  Set \m{requests} is (i) initialized to
empty, (ii) added to after receiving a \co{request} message, (iii) deleted
from under a condition about \m{decisions} in a \co{while} loop in the
top-level infinite loop, and (iv) added to under two nested conditions
inside a \co{while} loop after receiving a \co{decision} message.

\mypar{Higher-level executable specification}

To overcome the challenges in understanding the algorithm, we worked
hard %
to find the hidden properties and developed higher-level control flows and
synchronization conditions.

\begin{figure*}[htb!] %
\Hex{0}\begin{minipage}[c]{\arxiv{1.02}\ppdp{}\textwidth}
\rule{.99\columnwidth}{.1mm}\notppdp{\Vex{-1}}
\begin{code}
\ppdp{} 1 process Replica:
\ppdp{} 2   def setup(leaders, state): pass                                 # take in initial set of leaders and state
\ppdp{} 3   def run():
\ppdp{} 4     slot_in, slot_out := 1, 1                                     # slot to send prop, slot to apply decision
\ppdp{} 5     while true:
\ppdp{} 6       await slot_in < slot_out + WINDOW and                       # if slot_in can be increased and
\ppdp{} 7             some received ('request',c) has                       # some received request for command c
\ppdp{} 8               each sent ('propose',s,=c) has                      # each sent proposed slot s for c
\ppdp{} 9                 some received ('decision',=s,c2) has c2 != c:     # some received decison has s for a diff c2
\ppdp{}10         if some received ('decision', slot_in - WINDOW, (_,_,op)) has is_reconfig(op):
\ppdp{}11           leaders := op.leaders                                   # if slot_in-WINDOW is reconfig, set leaders
\ppdp{}12         if not some received ('decision',=slot_in,_):             # if slot_in is not decided
\ppdp{}13           send ('propose', slot_in, c) to leaders                 #   propose slot_in for command c
\ppdp{}14         slot_in := slot_in + 1\medskip
\ppdp{}15       or some received ('decision', =slot_out, c):                # if received decison slot_out for some c
\ppdp{}16         client, cmd_id, op := c                                   # extract components of command c
\ppdp{}17         if not (some received ('decision',s,=c) has s < slot_out) and not is_reconfig(op):
\ppdp{}18           state, result := apply(op, state)                       # if c not decided before \& is not reconfig
\ppdp{}19           send ('response', cmd_id, result) to client             #   apply op and send result to client
\ppdp{}20         slot_out := slot_out + 1\medskip
\ppdp{}21 process Leader:
\ppdp{}22   def setup(acceptors, replicas): pass                            # take in sets of acceptors and replicas
\ppdp{}23   def run():
\ppdp{}24     ballot := (0, self)                                           # ballot num is pair of round num and self
\ppdp{}25     while true:
\ppdp{}26       send ('1a', ballot) to acceptors                            # send 1a with ballot number           # 1a\medskip
\ppdp{}27       await count \{a: received ('1b',=ballot,_) from a\} > (count acceptors)/2:                           # 2a
\ppdp{}28         ps := \{p: received ('1b',=ballot,accepted), p in accepted\}           # all proposals accepted    #
\ppdp{}29         for (s,c) in \{(s,c): (b,s,c) in ps, b = max \{b2: (b2,=s,_) in ps\}\}:  # max b per s               #
\ppdp{}30           send ('2a', ballot, s, c) to acceptors                  # send 2a for previously accepted s,c  #
\ppdp{}31         while true:                                                                                      #
\ppdp{}32           await some received ('propose',s,c) has not some sent ('2a',=ballot,=s,_):                     #
\ppdp{}33             send ('2a', ballot, s, c) to acceptors                # send 2a for newly proposed s,c       #\medskip
\ppdp{}34           or some received ('2b',=ballot,s,c) has                                                   # learn
\ppdp{}35                count \{a: received ('2b',=ballot,=s,=c) from a\} > (count acceptors)/2:               #
\ppdp{}36             send ('decision', s, c) to replicas                   # send decided s,c to replicas    #\medskip
\ppdp{}37           or some received ('preempt',(r2,leader2)) has (r2,leader2) > ballot: break              # preempted
\ppdp{}38       or some received ('preempt',(r2,leader2)) has (r2,leader2) > ballot: pass                   # preempted\medskip
\ppdp{}39       ballot := (r2+1, self)                                      # increment round number in ballot number\medskip
\ppdp{}40 process Acceptor:
\ppdp{}41   def run(): await false                                          # wait for nothing, only to handle messages \medskip
\ppdp{}42   receive ('1a', b) from leader:                                  # Phase 1b: receive 1a with ballot b
\ppdp{}43     if each sent ('1b',b2,_) has b > b2:                          #  if b > each b2 in sent 1b msgs
\ppdp{}44       accepted := \{(b,s,c): sent ('2b',b,s,c)\}                    #  get accepted triples sent in 2b msgs
\ppdp{}45       send ('1b', b, accepted) to leader                          #  send 1b with b and accepted triples\medskip
\ppdp{}46   receive ('2a', b, s, c) from leader:                            # Phase 2b: receive 2a with bal,slot,cmd
\ppdp{}47     if not some sent ('1b',b2,_) has b2 > b:                      #  if not sent 1b with larger ballot b2
\ppdp{}48       send ('2b', b, s, c) to leader                              #  send 2b to accept the proposal triple\medskip
\ppdp{}49   receive m from leader:                                          # Preemption: receive 1a or 2a msg
\ppdp{}50     maxb := max (\{b: received ('1a',b)\} + \{b: received ('2a',b,_,_)\})  # find max bal in 1a,2a msgs
\ppdp{}51     if m[1] < maxb: send ('preempt', maxb) to leader              # if ballot in m < max bal, send preempt
\end{code}\Vex{-3}\ppdp{}
\rule{1\columnwidth}{.1mm}
\end{minipage}\noticdcs{\Vex{-1}}\icdcs{}

  \caption{A high-level specification of vRA Multi-Paxos in DistAlgo.\icdcs{}}
  \label{fig-vrapaxos-da}
\end{figure*}

Figure~\ref{fig-vrapaxos-da} shows a complete higher-level executable
specification of vRA Multi-Paxos in DistAlgo.
Even though executable code is generally much longer and more complex than
pseudocode, our specification is smaller than vRA Multi-Paxos pseudocode
(51 vs.\ 142 lines, or 100 lines without ``end'' of 42 scopes).  More
importantly, it is much simpler.
The new organization and main simplifications are as follows.  Other
improvements are described in Section~\ref{sec-fixes}, after understanding
the overall algorithm better in Section~\ref{sec-multi-understanding}.
\begin{description}
  \icdcs{}
  \setlength{\parskip}{.5ex}

\item {\bf Scout and Commander are removed.}  Their roles for Phases 1 and
  2 are merged into Leader.  Their roles for determining preemption are
  merged into Acceptor.

\item {\bf Repeated and scattered updates are replaced by high-level
    queries.}
  Leader collects majority from Phases 1b and 2b using two \co{count}
  queries (lines 27 and 35), not repeated updates of two sets (\m{\it
    waitfor}) in Scout and Commander; finds previously accepted and
  newly proposed proposals using only comprehensions and \co{some}
  queries (lines 28-29 and 32), not maintaining and updating a set
  (\m{\it proposals}); and confirms preemption using two \co{some}
  queries (lines 37 and 38), not maintaining an overall variable
  (\m{\it active}) and updating it (to \co{false} after premption and
  \co{true} elsewhere).

  Acceptor uses \co{each} and \co{some} (lines 43 and 47) and a comprehension
  (line 44), instead of using and maintaining a maximum (\m{\it
    ballot{\tiny\_}num}) and a set (\m{\it accepted}) by updates; and it
  determines preemption using a \co{max} query (line 50) in one place,
  instead of
  always sending \co{1b} and \co{2b} messages to scouts and commanders %
  and letting them determine preemption.

  Replica uses two clearly separated conditions, one for proposing
  commands based on requests, and one for applying commands based on
  decisions, instead of maintaining three sets (\m{requests},
  \m{proposals}, and \m{decisions}) by additions and deletions in
  mixed control flows and loop structures.

\end{description}\Vex{-1}

\mysubpar{High-level specification via declarative queries}
Examine the entire algorithm specification in Figure~\ref{fig-vrapaxos-da}.
\begin{itemize}
\item In process Replica for replicated state machines with
  reconfiguration, there are of course state variables updated for the
  application state (\co{slot\_in}, \co{slot\_out}, and \co{state}) and
  reconfiguration state (\co{leader}); they are orthogonal to the consensus
  algorithm run by Leader and Acceptor processes.  All other variables in
  assignments (\co{client}, \co{cmd\_id}, \co{op}, and \co{result}) are
  only temporary variables to be consumed immediately.
\item In processes Leader and Acceptor for Multi-Paxos with preemption,
  there is only one state variable, \co{ballot}, and it is repeatedly
  updated.  All other variables in assignments (\co{ps}, \co{accepted}, and
  \co{maxb}) are only temporary variables to be consumed immediately.
\end{itemize}

Updating state in replicated state machines for applications is of course
essential, just as sending and receiving messages are for distributed
algorithms.  However, the rest of the algorithm is specified almost
completely declarative, by using declarative queries over \co{sent} and
\co{received}.

The only exception is the update to \co{ballot}, which identifies successive
rounds, as discussed in Section~\ref{sec-basic-understanding}.   
This minimum state for tracking progress is fundamental in 
distributed systems, just like using logical clocks~\cite{Lam78}.

\mypar{Understanding Multi-Paxos with preemption and reconfiguration}
\label{sec-multi-understanding}

Without low-level updates, we now see how Figure~\ref{fig-vrapaxos-da} 
precisely extends Basic Paxos with continuous slots, preemption, 
replicated state machine, and reconfiguration.

A Leader process takes a set \co{acceptors} and a set \co{replicas} (line
22), initializes \co{ballot} to round 0 paired with \co{self} (line 24),
and repeatedly (line 25) does two things with each incremented \co{ballot}
(line 39).
\begin{enumerate}

\item[1.] Send \co{1a} message for the current \co{ballot} to \co{acceptors}
  (line 26) and wait for a majority \co{1b} replies for the \co{ballot}
  (line 27), as in Proposer in Phase 1 of Basic Paxos; but also wait for
  some \co{preempt} with a larger ballot \co{(r2,leader2)} than the
  \co{ballot} and then do nothing more for the current \co{ballot} (line
  38).

  Note that between receiving a majority \co{1b} messages and receiving a
  \co{preempt} is exactly when the original pseudocode maintains
  variable \co{active} as \co{true} for the current \co{ballot}.  

\item[2.] After receiving a majority \co{1b} replies, (1) find in \co{1b}
  messages (line 28) previously accepted slot \co{s} and command \co{c}
  pairs that correspond to the largest ballot for each \co{s} (line 29) and
  send them with the current \co{ballot} to \co{acceptors} (line 30),
  ensuring that they %
  continue to be accepted in the current ballot, and (2) repeatedly (line
  31) wait to (i) receive newly proposed \co{s} and \co{c} for which no
  \co{2a} message has been sent for \co{s} for the \co{ballot} (line 32)
  and send a \co{2a} message for \co{s} and \co{c} to \co{acceptors} (line
  33), or (ii) receive a majority \co{2b} replies for some \co{s} and
  \co{c} (lines 34-35) and send \co{s} and \co{c} as a decision to
  \co{replicas} (line 36), or (iii) receive a \co{preempt} with a larger
  ballot than the \co{ballot} and break out of the repeats.

  This is as in Proposer in Phase 2 of Basic Paxos except that a \co{1b}
  message holds a set of triples
  instead of a single pair; a \co{2a} message is sent for a
  triple %
  (\co{ballot}, \co{s}, \co{c}) for each slot \co{s}
  (where (\co{s}, \co{c}) pairs in received \co{propose} messages are
  the allowed values) instead of a single
  pair %
  (\co{n}, \co{v}) (where \co{v} in 1..100 are allowed values); 
  a \co{2b} message has an additional slot component too,
  and is received by the leader instead of a learner, and a \co{decision}
  is sent back to \co{replicas} instead of a \co{chosen} being
  outputted; and receiving a \co{preempt} message is added.

\end{enumerate}

An Acceptor process continuously waits (line 41) to receive \co{1a} and
\co{2a} messages and does one of two things:
\begin{itemize}

\item Use \co{each} and \co{some} queries to check sent \co{1b} messages
  (lines 43 and 47), and send back \co{1b} and \co{2b} messages (lines 45
  and 48), exactly as in Acceptor in Phases 1 and 2 of Basic Paxos except
  with the additional \co{s} component in \co{2a} and \co{2b} messages, and
  computing a set \co{accepted} of proposals instead of a single
  \co{max\_prop} proposal.

\item Test if the ballot in a received \co{1a} or \co{2a} message is less
  than the maximum ballot ever received (line 50), and send back
  \co{preempt} with the maximum ballot (line 51).
  This is not only simpler and more direct, it is also much more efficient
  than always sending \co{1b} messages with the set \co{accepted} that is
  large and will be ignored anyway.
\end{itemize}

A Replica process takes a set \co{leaders} and a \co{state} (line 2),
initializes \co{slot\_in} and \co{slot\_out}
to 1 (line 4), and repeatedly (line 5) executes one of two branches and
increments \co{slot\_in} or \co{slot\_out} (lines 14 and 20):
\begin{itemize}

\item If some received requested command \co{c} (line 7) %
  is such %
  that each sent proposed slot \co{s} for \co{c} (line 8) has \co{s} taken
  by a different command \co{c2} in some received decision (line 9),
  propose \co{slot\_in} for \co{c} to \co{leaders} (line 13) if
  \co{slot\_in} is not already used in some received decision (line 12).
  For reconfiguration, check also that \co{slot\_in} is within \co{WINDOW}
  slots ahead of \co{slot\_out} (line 6), and if the decision at
  \co{WINDOW} slots back is a command for a reconfiguration operation (line
  10), set new leaders to be those in that operation (line 11).

\item If some received decided command \co{c} is for \co{slot\_out} (line
  15), get the client, command id, and operation in \co{c} (line 16), and
  if there was not already a decision for \co{c} in some earlier slot
  \co{s} and the operation is not reconfiguration (line 17), apply the
  operation to \co{state} to obtain a new state and a result (line 18), and
  send the result with the command id to the client (line 19).

\end{itemize}

The complete preemption functionality is expressed simply on lines 50-51
and 37-38, and reconfiguration is completely expressed on lines 6 and 10-11
with two increments on lines 14 and 20.  These are easy to see, again due
to our use of high-level queries for control flows and synchronization
conditions.

\mysec{Issues and fixes}
\label{sec-fixes}

This section describes useless replies, unnecessary delays, liveness
violations and fixes discovered in developing the specification in
Figure~\ref{fig-vrapaxos-da}.
All problems to be described were difficult to find in the original
pseudocode due to complex control flows.  Developing higher-level
specifications, especially 
using nondeterministic \co{await} with
message-history queries for
synchronization conditions, 
helped us understand the algorithm better and
discover these problems easily, by just following the simplified
algorithm flows.

The useless replies and unnecessary delays are small inefficiencies, 
but they violated invariants, making it hard to understand the pseudocode,
find the invariants, and write high-level queries.  
The liveness violations %
are easy to find after high-level queries are used.

The liveness violations can occur if messages can be lost.  The liveness
violation in Replica was confirmed by author 
van Renesse when we first discovered it.  However, we recalled later that
the paper has used a fair links assumption.  It is a strong assumption,
and assumes that messages are
periodically retransmitted until an ack is received.  Nonetheless,
implementing such retransmission is not easy without slowing down the
executions too much, because even TCP connections can break, and
retransmission would need to include repairing broken TCP connections.  The
Ovid framework by the authors~\cite{ovid2016} does such repair and
retransmission automatically.  When we discovered the liveness violation in
Leader, we learned from author van Renesse that making a framework like
Ovid efficient is very difficult.  We describe fixes without relying on 
message retransmission.

\begin{description}

\item {\bf Useless replies in Acceptor are fixed.}
In %
Figure~\ref{fig-vrapaxos-da}, if an acceptor receives a \co{2a} message
with a larger ballot \co{b} than the maximum ballot in all sent \co{1b}
messages, i.e., all received \co{1a} messages, it replies with the same
ballot \co{b} (line 48), which will be used when counting majority for
ballot \co{b}, exactly as in Basic Paxos.  In the original pseudocode, it
replies with that maximum (line 15 in Figure~4 %
in~\cite{vra15paxos}), smaller than \co{b}, causing the reply to be ignored by
the Commander and Leader processes, rendering the reply useless.

\item {\bf Unnecessary delays in Replica are fixed.}
In Figure~\ref{fig-vrapaxos-da}, the first \co{await} condition (lines 6-9)
allows any received request, for which each sent proposal for a slot has
that slot taken by a different command in a received decision, to be
detected and re-proposed immediately.  The original pseudocode delays the
detection and re-proposal until \co{slot\_out} equals the taken slot.

These delays could be relatively minor by themselves, but realizing them and
removing them helped us develop our simpler specification, especially in
terms of control flows, which subsequently led us to easily discover the
liveness violation in Replica.

\item {\bf Liveness violation in Replica and fix.}
Replica is arguably the most complex of the process types: it needs to
mediate with both clients and leaders while competing with other replicas
for proposed slots; perform eventual state updates in correct order of
decided slots despite possibly receiving them out of order; and
support reconfiguration. %

Our high-level specification led us to discover a liveness violation in the
original pseudocode: if no decision is received for a slot, e.g., due to lost
\co{propose} messages from all replicas proposing for that slot,
all replicas will stop applying decisions from that slot on, so
\co{slot\_out} will stop incrementing; furthermore, due to the limited
\co{WINDOW} used for incrementing \co{slot\_in}, all sending of proposals
will stop after the \co{WINDOW} is used up.  So all replicas will be
completely stuck, and the entire system will stop making progress.

To fix, a replica can propose for that slot again after a timeout.  A
leader can then work on deciding for that slot if a decision for it has not
been made; otherwise, it can send back the decision for that slot.

\item {\bf Liveness violation in Leader and fix.}
Leader is the core of Paxos.  
A leader must always be able to make progress unless it has crashed.
However, liveness violations can occur in several ways.

For Phase 1, if after sending a \co{1a} message, the leader does not
receive \co{1b} messages from a majority of acceptors or a \co{preempt}
message, e.g., due to a lost \co{1a} message, the leader will wait forever.
To fix, the leader can add a timeout to the outer \co{await}, so a new
round of Phase 1 will be started after the timeout.

For Phase 2, if for a newly proposed slot, no \co{2b} messages are received
from a majority, e.g., due to a lost \co{2b} message, the leader will not
make a decision for that slot.  If this happens to all leaders, the
replicas will not receive a decision for that slot, leading to the liveness
violation in Replica described earlier.
To fix, the leader can send the \co{2a} message with that slot again after
a timeout from waiting to receive a majority of \co{2b} messages for it.

For Phase 2, if a \co{preempt} message is lost, and a majority has seen a
larger ballot, the leader will fruitlessly continue to send \co{2a}
messages for newly proposed \co{s,c} and count received \co{2b} messages 
but not having a majority to make any \co{decision}, and thus not making
progress.
To fix, the leader can start a new round of Phase 1 after a timeout from
waiting to receive a majority or a preemption.

\end{description}
\icdcs{}%

\mysec{%
Optimizations and executions}
\label{sec-optimize}

High-level specifications can be too inefficient to execute, which is
indeed the case with the specification in Figure~\ref{fig-vrapaxos-da} as
well as any specification following the original
pseudocode~\cite{vra15paxos}.
However, high-level specifications also allow additional optimizations and
extensions to be done more easily.  We specify the two most important ones
suggested in~\cite{vra15paxos} for vRA Multi-Paxos but not included in its
pseudocode, describe a general method for merging processes that supports a
range of additional optimizations, and discuss results of executions.

\mypar{State reduction with maximum ballot}
\label{sec-state-red}

The most serious efficiency problem of the algorithm in
Figure~\ref{fig-vrapaxos-da} is the fast growing set \co{accepted} in \co{1b} messages, 
which quickly chokes any execution of the algorithm that does real message
passing.
The solution is to not keep all triples in sent \co{2b} messages, as in
Figure~\ref{fig-vrapaxos-da} and the original pseudocode~\cite{vra15paxos},
but keep only triples with the maximum ballot for each slot, so there is at
most one triple for each slot.  This is done by changing line 44 to
\arxiv{\Vex{1}}
\begin{code}
\arxiv{    44       accepted := \{(b,s,c): sent ('2b',b,s,c), b = max \{b: sent ('2b',b,=s,_)\}\}}\twocol{}
\end{code}
\ppdp{}
This drastically reduces not only the size of \co{1b} messages, but also the space needed by acceptors and leaders, which send and receive \co{1b} messages, respectively.

\mypar{Failure detection with ping-pong and timeout}
\label{sec-fail-detect}

Failure detection addresses the next most serious problem: leaders compete
unnecessarily to become the leader with the highest ballot, leaving little
or no time for proposals to be decided.
Adding failure detection uses ping-pong after preemption: in \co{Leader},
after exiting the outer \co{await} following a \co{preempt} and before
incrementing \co{ballot},
periodically ping the leader \co{leader2} that has the larger ballot
\co{(r2,leader2)} and wait for replies, by inserting
\ppdp{}
{
\begin{code}
\arxiv{    38.1       while each sent('ping',=r2,t) to =leader2 has received ('pong',r2,t) from leader2:
    38.2         send ('ping', r2, logical_time()) to leader2
    38.3         await timeout TIMEOUT}\twocol{}
\end{code}
}
\ppdp{}
\noindent 
and adding the following \co{receive} definition after the \co{run} definition:
\ppdp{}
{
\begin{code}
\arxiv{    }39.1   receive ('ping', r2, t) from leader2:
\arxiv{    }39.2     send ('pong', r2, t) to leader2 
\end{code}
}
\ppdp{}
\noindent
\co{TIMEOUT} is a variable holding the timeout value in seconds.

\mypar{Merging processes}

High-level specifications in DistAlgo allow different types of processes
that run at the same time to be merged easily, even if they interact with
each other in sophisticated ways, provided they together have one main flow
of control.  There are two cases:
\begin{enumerate}\lipics{}

\item[1.] A process \co{\p{P}} that has only \co{await false} in \co{run} can be
  merged easily with any process \co{\p{Q}}.  For example, 
  in Figure~\ref{fig-vrapaxos-da}, \co{Acceptor} can be merged
  with \co{Leader} by adding the \co{receive} definitions of \co{Acceptor} to
  the body of \co{Leader}.

\item[2.] A process \co{\p{P}} that has only \co{while true:~await...} in
  \co{run}, with no \co{timeout} in the \co{await}, can be merged easily with
  any process \co{\p{Q}}.  For example, %
  in Figure~\ref{fig-vrapaxos-da}, \co{Replica} can be merged with
  \co{Leader} by adding, for each branch \co{\p{cond}:\,\p{stmt}} of
  \co{await} of \co{Replica}, a \co{receive \_:~if \p{cond}:\,\p{stmt}}
  definition to the body of \co{Leader}.

\end{enumerate}\lipics{}
Process setups can be transformed accordingly.  Details are omitted because
they are less important.
These transformations are easy to automate.  Inversely, independent
\co{receive} definitions can be easily put into separate processes.
In Figure~\ref{fig-vrapaxos-da}, all three types of
processes, or any two of them, can be merged, giving a total of 4 possible
merged specifications.

Merging supports colocation of processes cleanly, and allows a range of
optimizations, e.g., %
garbage collection of states of leaders and acceptors, for decided slots
already learned by all replicas~\cite{vra15paxos}.
Furthermore, communication between processes that are merged no longer
needs real message passing but can be done more efficiently through shared
memory.  Also, because the actions are independent, lightweight threads can
be used to make each process more efficient.

With a few more small variations to vRA Multi-Paxos, merging Replica,
Leader, and Acceptor into one process type yields essentially the 'Replica'
in Chubby~\cite{burrows06chubby}, Google's distributed lock service that
uses Paxos, and the 'Server' in Raft~\cite{ongaro14raft}, a pseudocode for
the main features of Chubby.  In general, separate processes provide
modularity, and merged processes reduce overhead.  Being able to merge
separate processes easily allows one to obtain the benefits of both.

\mypar{Configuration and execution}

A \co{main} definition, similar to that in Figure~\ref{fig-lapaxos-da}, can
set up a number of \co{Replica}, \co{Leader}, and \co{Acceptor} processes,
or their merged versions, and some \co{Client} processes that send
\co{request} messages to \co{Replica} processes and receive \co{response}
messages; and parameters \co{WINDOW} and \co{TIMEOUT} can be defined.
We summarize the results of running DistAlgo specifications:
\begin{itemize}

\item DistAlgo specifications can be run directly. For example, a complete
  specification of vRA Multi-Paxos with state reduction and failure
  detection in file \co{spec.da} (available at
  \anonymous{}%
  \notanonymous{\url{darlab.cs.stonybrook.edu/paxos}})
  can be run %
  by executing %
  \co{pip install pyDistAlgo} followed by 
  \co{python -m da spec.da}.
  The default is to run all processes on the local machine as separate
  operating system processes.

\item The DistAlgo specification for vRA Multi-Paxos as in
  Figure~\ref{fig-vrapaxos-da}, without the state reduction in
  Section~\ref{sec-state-red}, almost immediately overflows the default
  message buffer size of 4KB, yielding a \co{MessageTooBigException}.

\item The DistAlgo specification for vRA Multi-Paxos with state reduction,
  without the failure detection in Section~\ref{sec-fail-detect}, runs
  continuously but most times stops making progress (decisions) for 3
  leaders, 3 acceptors, and 3 replicas, serving 10 client requests, and was
  killed manually after 200 rounds (200--600 ballots) have been attempted.

\item The DistAlgo specification for vRA Multi-Paxos with both state
  reduction and failure detection runs smoothly.  For example, for %
  10 processes (3 leaders, 3 acceptors, 3 replicas, and 1 client), %
  processing 10 requests takes 77.822 %
  milliseconds (ranging from 74.141 %
  to 84.825), %
  averaged over 10 runs, on a %
  Intel Core i7-6650U 2.20GHz CPU %
  with 16 GB RAM, running Windows 10 and Python 3.7.0. %

\end{itemize}

\mysubpar{Additional optimizations}
Many additional optimizations and experiments %
can be done, especially including transforming high-level queries into
efficient 
incremental updates of auxiliary variables~\cite{Liu+12DistPL-OOPSLA,Liu+17DistPL-TOPLAS}, 
but they are beyond the scope of this paper.
Note that incremental updates of auxiliary variables
allow \co{sent} and \co{received} to become dead variables and be eliminated.
Our experience is that precise high-level specifications %
allow us to understand the algorithms much better and to significantly
improve both correctness and efficiency much more easily than was possible
before.

\noticdcs{
\mysec{Correctness and formal verification}
\label{sec-verify}

The problems of vRA Multi-Paxos described in Section~\ref{sec-fixes} do not
affect the safety of vRA Multi-Paxos.  However, even safety is not easy to
understand and needs verification.

We have developed formal proofs of safety of the complete specification of
vRA Multi-Paxos in Figure~\ref{fig-vrapaxos-da}, and of the one extended
with state reduction and the one further extended with failure detection as
described in Section~\ref{sec-optimize}.
The safety property ensures that, for each slot, only a single command may
be decided and it must be one of the commands proposed.

The proofs are done by first translating the specifications into TLA+,
Lamport's Temporal Logic of Actions~\cite{Lam02book}.  This was done
systematically but manually.
The high-level nature of our specifications makes the translation simple 
conceptually: each type of data in DistAlgo corresponds to a type of data
in TLA+, and each expression and statement in DistAlgo corresponds to a
conjunction of equations in TLA+.
The three translated specifications of vRA Multi-Paxos and extensions in
TLA+ are 154, 157, and 217 lines.

Automatic translators from DistAlgo to TLA+ had in fact been developed
previously, and we started developing a new one~\cite{Liu+8DistInv-TLA}.  
The earlier
translators produced TLA+ specifications that contain many more details
needed to handle general control flows, especially low-level flows, and are
formidable for formal verification, even for simpler protocols.
We are leveraging experience from systematic manual translations of high-level
specifications in building a new automatic translator.

\mypar{Proofs in TLAPS}

Our proofs are manually developed and automatically machine-checked using
TLAPS~\cite{tlaps15}, a proof system for TLA+.  The proofs for the three
translated specifications are 4959, 5005, and 7006 lines, respectively.
The proofs are much more complex and longer than the previous proof of 1033
lines for Multi-Paxos with preemption~\cite{Cha+16PaxosTLAPS-FM}, because
of the additional details in vRA Multi-Paxos, and the extensions for state
reduction and failure detection.
Appendix~\ref{app-proofs} contains additional details about the proofs.

\notes{} %

Compared to %
other manually developed and mechanically checked proofs
of executable Multi-Paxos and variants, namely, a proof of safety and
liveness of Multi-Paxos in Dafny from
IronFleet~\cite{hawblitzel2015ironfleet} and a proof of safety of
Raft~\cite{ongaro14raft} in Coq from Verdi~\cite{woos2016planning}, %
our proofs in TLAPS are %
4 to 10 times smaller, %
compared with 30,000 lines in IronFleet and 50,000 lines in Verdi.
We believe this is due to our use of high-level queries over message
histories, which are much simpler and directly capture important
invariants, in contrast to use of repeated and scattered lower-level
updates, as studied for a more abstract
specification~\cite{ChaLiu18PaxosHistVarTLAPS-NFM}.
Shorter proofs are much easier to understand and maintain, and also easier
and faster to check automatically.  Both are significant advantages for
practical development cycles of specifications, programs, and proofs.

Our proof checking times are 9.5, 9.8, and 13 minutes, respectively, on an
Intel i7-4720HQ 2.6GHz CPU with 16GB memory running Ubuntu16.04 LTS and
TLAPS1.5.2.
No proof checking time is reported for the proof from
Verdi~\cite{woos2016planning}, but we were able to run proof check for the
proof after solving some version mismatch problems, and it took 29 minutes
to run on the same machine as our proof.
The proof checking times for the proofs from IronFleet are reported to be
147 minutes for the protocol-level proof and 312 minutes including also the
implementation-level proof (without specifying the machine used for the
proofs)~\cite{hawblitzel2015ironfleet}; we have not been able to run proof
check for their proofs on our machine due to an error from a build file.

\mypar{Safety violation and fix}
\label{sec-safety-violation}

Our development of formal proofs also allowed us to discover and fix a
safety violation in an earlier version of our specification for vRA
Multi-Paxos.  There, acceptors always reply with \co{1b} and \co{2b}
messages, not \co{preempt} messages, as in the original pseudocode, and
leaders try to detect preemption.
It is incorrect because the ballot number in leaders may increase after a
\co{1a} or \co{2a} message is sent, contrasting the fixed ballot number in
a Scout or Commander process used for detecting preemption.
The safety violation was discovered after the proof could not succeed.

The fix of having acceptors detect preemption and inform the leader also
makes the algorithm much more efficient in the case of preemption upon
receiving \co{1a} messages: a \co{preempt} message with only a ballot
number is sent, as in Figure~\ref{fig-vrapaxos-da}, instead of a \co{1b}
message with a ballot number and an entire \co{accepted} set, as in the
original pseudocode.

The earlier incorrect version was used in distributed algorithms and
distributed systems courses for several years,
with dozens of course projects and homeworks having used it, including ones
directed specifically at testing and even modeling using TLA+ and model
checking using TLC~\cite{tlatoolbox18}.
However, this safety violation was never found, because it requires delays
of many messages, extremely unlikely to be found by testing or model
checking, due to the large search space that must be explored.
} %

\mysec{Related work and conclusion}
\label{sec-related}

Consensus algorithms and variants around Paxos have seen a long series of
studies, especially their specifications for understanding, implementation,
and verification.

Paxos is well known to be hard to understand.  Since its initial
description~\cite{Lam98paxos}, much effort has been devoted to its better
exposition and understanding.
Earlier descriptions use English, e.g.,~\cite{lam01paxos}, or state
machines, e.g.,~\cite{lampson1996build,prisco00revisit}.  Later studies
include pseudocode, e.g.,~\cite{kirsch2008paxosTR,ongaro14raft,vra15paxos},
and deconstructed pseudocode or code,
e.g.,~\cite{boichat2003deconstructing,van15vive,garcia18paxos}.
Among existing works, vRA Multi-Paxos pseudocode~\cite{vra15paxos} is by
far the most direct, complete, and concise specification of a more
realistic version of Multi-Paxos.
Our specification captures and improves over vRA Multi-Paxos pseudocode,
and yet is much simpler and smaller, even though executable code is generally
much larger and more complex than pseudocode.

In particular, our specification captures the control flows and 
synchronization conditions
at a higher level than previous specifications, yet is precise, complete,
and directly executable.  It is exactly the higher-level, simpler
specification
that allowed us to easily
discover the liveness violations in vRA Multi-Paxos if 
messages can be lost, and to find other issues and fixes.
It has also helped tremendously in teaching~\cite{Liu+17DistPL-TOPLAS}.

An earlier work~\cite{Liu+12DistSpec-SSS} uses similar high-level language
constructs.  However, it keeps the complex Leader process doing low-level
updates while spawning Scout and Commander processes as in vRA Multi-Paxos.
It does not have reconfiguration, state reduction, or failure detection;
its more complex control flows make it more difficult to add them and be
sure of correctness.
Its Replica process uses a \co{for} loop and sequential statements in the
loop body instead of high-level \co{await} conditions, causing it to
iterate extremely inefficiently, bias towards sending a proposal before
applying decisions, and apply decisions exhaustively before sending more
proposals.  It also has the same liveness violations as in vRA Multi-Paxos
but they were not discovered.

One might also try to understand Paxos variants through more practical
implementations, but these implementations %
are much larger and more complex. %
For example, Google Chubby's C++ server code is reported to be about 7000
lines~\cite{burrows06chubby,chandra07paxos} and is not open-source.
Paxos for system builders~\cite{kirsch2008paxos,kirsch2008paxosTR} is
written in C and has over 5000 lines, not including over 2000 lines of
library code for group communication primitives.\footnote{From the
  ``Download'' link of ``Paxos for System Builders'' at
  \url{http://www.dsn.jhu.edu/software.html}, April 15, 2018.}
OpenReplica~\cite{altinbuken2012commodifying} is implemented in Python and
has about 3000 lines.\footnote{From email with Emin Gun Sirer, August 12,
  2011.}
They generally include many lower-level data structures, bookkeeping tasks,
and language details.
This makes it much harder to understand the algorithms used.  They are also
so far infeasible for formal verification, manually or with automated
support.
A more feasible approach is to generate low-level data structure 
updates from high-level specifications, e.g., as studied
in~\cite{Liu+12DistPL-OOPSLA,Liu+17DistPL-TOPLAS}.

\notes{} %

There has been significant effort on formal specification and verification
of Paxos and variants.
Many specifications have been developed, %
especially for Basic Paxos, including earlier ones described
previously~\cite{Liu+12DistPL-OOPSLA}; our specifications are significantly
simpler while including full details for execution in real distributed
environments.
Several proofs are successful.
Some are automatic by writing specifications %
in a restricted language~\cite{padon2017paxos,taube2018modular}.
Some include proofs of liveness properties,
e.g.,~\cite{hawblitzel2015ironfleet,padon2017reducing}.
Some also generate code, e.g.,~\cite{georgiou09automated,taube2018modular}.
Some proofs are discussed extensively but are only on
paper~\cite{garcia18paxos}.
Some others are also for abstract specifications that omit many algorithm
details,
e.g.,~\cite{Cha+16PaxosTLAPS-FM,bobba2017design,ChaLiu18PaxosHistVarTLAPS-NFM}.
Proofs for executable implementations are from
IronFleet~\cite{hawblitzel2015ironfleet} and Verdi~\cite{wilcox2015verdi},
with much larger proofs and longer proof checking times\noticdcs{, as discussed in 
Section~\ref{sec-verify}}.

To the best of our knowledge, no previous efforts of specification and
formal verification, for any Paxos variant, reported finding any
correctness violations in published specifications or any improvements to
them.  However, Fonseca et al.~\cite{fonseca2017empirical} discovered 16
bugs in IronFleet, Verdi, and Chapar~\cite{lesani2016chapar} for
distributed key-value stores.  
These include bugs in protocol specification, verification tool, and shim
layer modeling;
no bugs were found in the protocols modeled.  
11 of the 16 bugs were discovered by manual inspection, even without prior
experience of Fonseca with OCaml, the language used by Coq.\footnote{From
  communication with %
  Fonseca at and after a talk he gave, April 2018.} %
This helps support both the importance and difficulty of writing
good specifications for understanding and manual inspection, as well as
formal verification.

\notes{}

There are many directions for future research: higher-level specifications
of more variants of Paxos and other important distributed algorithms, %
more powerful optimizations for automatically generating efficient
implementations, and better methods for developing automated proofs
directly from high-level specifications.

\noticdcs{
\appendix

\mysec{van Renesse and Altinbuken's pseudocode for Multi-Paxos with
  preemption and reconfiguration}
\label{app-vrapaxos}

Figure~\ref{fig-vra-leader} shows the Leader process that spawns Scout and Commander
processes in vRA Multi-Paxos.  Replica and Acceptor processes are not shown
due to space limitations.

%%%\documentclass{article}[10pt]

%%%\usepackage{xspace}
%%%\usepackage{fullpage}

%%%\begin{document}

% indentation in pseudocode
\newcommand{\ind}{\hspace*{0.7em}}

% math constructs
\newcommand{\tuple}[1]{\langle #1\rangle}
\newcommand{\setc}[2]{\{#1 \;|\; #2\}}

% variables
\newcommand{\vstate}{{\it state}\xspace}
\newcommand{\vslotin}{{\it slot{\tiny\_}in}\xspace}
\newcommand{\vslotout}{{\it slot{\tiny\_}out}\xspace}
\newcommand{\vrequests}{{\it requests}\xspace}
\newcommand{\vproposals}{{\it proposals}\xspace}
\newcommand{\vdecisions}{{\it decisions}\xspace}
\newcommand{\vop}{{\it op}\xspace}
\newcommand{\vleaders}{{\it leaders}\xspace}
\newcommand{\vcid}{{\it cid}\xspace}
\newcommand{\vacceptors}{{\it acceptors}\xspace}
\newcommand{\vpvalues}{{\it pvalues}\xspace}
\newcommand{\vwaitfor}{{\it waitfor}\xspace}
\newcommand{\vreplicas}{{\it replicas}\xspace}
\newcommand{\vactive}{{\it active}\xspace}
\newcommand{\vballotnum}{{\it ballot{\tiny\_}num}\xspace}
\newcommand{\vself}{{\it self}()\xspace}
\newcommand{\vpvals}{{\it pvals}\xspace}
\newcommand{\vaccepted}{{\it accepted}\xspace}

% constants
\newcommand{\cwindow}{\mbox{WINDOW}\xspace}

\small

%end notes

\begin{figure*}[htbp]
\arxiv{\scriptsize}
\ppdp{\small}
  \centering
\begin{center}
\fbox{
\begin{tabular}[t]{@{}l@{}}
process {\it Scout}($\lambda, \vacceptors, b$)\\
\ind var $\vwaitfor := \vacceptors$, $\vpvalues := \emptyset$;\smallskip\\
\ind $\forall \alpha \in \vacceptors : {\it send}(\alpha, \tuple{{\bf p1a}, \vself, b})$;\\
\ind for ever\\
\ind\ind switch {\it receive}()\\
\ind\ind\ind case $\tuple{{\bf p1b}, \alpha, b', r}$ :\\
\ind\ind\ind\ind if $b' = b$ then\\
\ind\ind\ind\ind\ind $\vpvalues := \vpvalues \cup r$;\\
\ind\ind\ind\ind\ind $\vwaitfor := \vwaitfor \setminus \{\alpha\}$;\\
\ind\ind\ind\ind\ind if $\vert \vwaitfor\vert < \vert \vacceptors\vert/2$ then\\
\ind\ind\ind\ind\ind\ind {\it send}($\lambda, \tuple{{\bf adopted}, b, \vpvalues}$);\\
\ind\ind\ind\ind\ind\ind {\it exit}();\\
\ind\ind\ind\ind\ind end if\\
\ind\ind\ind\ind else\\
\ind\ind\ind\ind\ind {\it send}($\lambda, \tuple{{\bf preempted}, b'}$);\\
\ind\ind\ind\ind\ind {\it exit}();\\
\ind\ind\ind\ind end if\\
\ind\ind\ind end case\\
\ind\ind end switch\\
\ind end for\\
end process
\end{tabular}}
%\end{center}
\Hex{.1}
%\begin{center}
\fbox{
\begin{tabular}[t]{@{}l@{}}
process {\it Commander}($\lambda, \vacceptors, \vreplicas, \tuple{b, s, c}$)\\
\ind var $\vwaitfor := \vacceptors$;\smallskip\\
\ind $\forall \alpha \in \vacceptors : {\it send}(\alpha, \tuple{{\bf p2a}, \vself, \tuple{b, s, c}})$;\\
\ind for ever\\
\ind\ind switch {\it receive}()\\
\ind\ind\ind case $\tuple{{\bf p2b}, \alpha, b'}$ :\\
\ind\ind\ind\ind if $b' = b$ then\\
\ind\ind\ind\ind\ind $\vwaitfor := \vwaitfor \setminus \{\alpha\}$;\\
\ind\ind\ind\ind\ind if $\vert \vwaitfor\vert < \vert \vacceptors\vert/2$ then\\
\ind\ind\ind\ind\ind\ind $\forall \rho \in \vreplicas :$\\
\ind\ind\ind\ind\ind\ind\ind {\it send}($\rho, \tuple{{\bf decision}, s, c}$);\\
\ind\ind\ind\ind\ind\ind {\it exit}();\\
\ind\ind\ind\ind\ind end if\\
\ind\ind\ind\ind else\\
\ind\ind\ind\ind\ind {\it send}($\lambda, \tuple{{\bf preempted}, b'}$);\\
\ind\ind\ind\ind\ind {\it exit}();\\
\ind\ind\ind\ind end if\\
\ind\ind\ind end case\\
\ind\ind end switch\\
\ind end for\\
end process
\end{tabular}}
\end{center}

\begin{center}
\fbox{
\begin{tabular}[t]{@{}l@{}}
process {\it Leader}($\vacceptors, \vreplicas$)\\
\ind var $\vballotnum := (0, \vself)$, $\vactive := \mbox{false}$, $\vproposals := \emptyset$;\smallskip\\

\ind {\it spawn}({\it Scout}($\vself, \vacceptors, \vballotnum)$);\\
\ind for ever\\
\ind\ind switch {\it receive}()\\
\ind\ind\ind case $\tuple{{\bf propose}, s, c}$ :\\
\ind\ind\ind\ind if $\not\exists c' : \tuple{s,c'} \in \vproposals$ then\\
\ind\ind\ind\ind\ind $\vproposals := \vproposals \cup \{\tuple{s,c}\}$;\\
\ind\ind\ind\ind\ind if \vactive then\\
\ind\ind\ind\ind\ind\ind {\it spawn}({\it Commander}($\vself, \vacceptors, \vreplicas, \tuple{\vballotnum, s, c}$));\\
\ind\ind\ind\ind\ind end if\\
\ind\ind\ind\ind end if\\
\ind\ind\ind end case\\
\ind\ind\ind case $\tuple{{\bf adopted}, \vballotnum, \vpvals}$ :\\
\ind\ind\ind\ind $\vproposals :=  \vproposals \triangleleft {\it pmax}(\vpvals)$;\\
\ind\ind\ind\ind $\forall \tuple{s,c} \in \vproposals$ :\\
\ind\ind\ind\ind\ind {\it spawn}({\it Commander}($\vself, \vacceptors, \vreplicas, \tuple{\vballotnum, s, c}$));\\
\ind\ind\ind\ind $\vactive := \mbox{true}$;\\
\ind\ind\ind end case\\
\ind\ind\ind case $\tuple{{\bf preempted}, \tuple{r', \lambda'}}$ :\\
\ind\ind\ind\ind if $(r', \lambda') > \vballotnum$ then\\
\ind\ind\ind\ind\ind $\vactive := \mbox{false}$;\\
\ind\ind\ind\ind\ind $\vballotnum := (r' + 1, \vself)$;\\
\ind\ind\ind\ind\ind {\it spawn}({\it Scout}($\vself, \vacceptors, \vballotnum$));\\
\ind\ind\ind\ind end if\\
\ind\ind\ind end case\\
\ind\ind end switch\\
\ind end for\\
end process
\end{tabular}}
\end{center}
%\Vex{-19}
\begin{center}
%\Hex{20}
\fbox{
$pmax\langle pvals\rangle \equiv 
\begin{array}[t]{@{}l}
  \{\langle s,c\rangle|\exists b:\langle b,s,c\rangle \in pvals\\
  \land \forall b',c':\langle b',s,c'\rangle \in pvals
  \Rightarrow b'\leq b\}
\end{array}$}\\
\vspace{.5ex}
%\Hex{20}
\fbox{
$x \triangleleft y \equiv
\begin{array}[t]{@{}l}
  \{\langle s,c\rangle|\langle s,c\rangle \in y\\
  {}\lor (\langle s,c\rangle \in x \land \not\exists c':\langle s,c'\rangle
  \in y)\}
\end{array}$}
\end{center}\Vex{-2}
\caption{Leader with Scout and Commander processes in vRA Multi-Paxos
  pseudocode~\cite[Fig.\,6, Fig.\,7, and the two definitions on pages
  12-14]{vra15paxos}.}
  \label{fig-vra-leader}
\end{figure*}

\notes{
\begin{figure}[htbp]
\small
  \centering
\begin{center}
\fbox{
\begin{tabular}[t]{@{}l@{}}
process {\it Acceptor}()\\
\ind var $\vballotnum := \bot$, $\vaccepted := \emptyset$;\smallskip\\

\ind for ever\\
\ind\ind switch {\it receive}()\\
\ind\ind\ind case $\tuple{{\bf p1a}, \lambda, b}$ :\\
\ind\ind\ind\ind if $b > \vballotnum$ then\\
\ind\ind\ind\ind\ind $\vballotnum := b$;\\
\ind\ind\ind\ind end if\\
\ind\ind\ind\ind {\it send}($\lambda, \tuple{{\bf p1b}, \vself, \vballotnum, \vaccepted}$);\\
\ind\ind\ind end case\\
\ind\ind\ind case $\tuple{{\bf p2a}, \lambda, \tuple{b, s, c}}$ :\\
\ind\ind\ind\ind if $b = \vballotnum$ then\\
\ind\ind\ind\ind\ind $\vaccepted := \vaccepted \cup \{\tuple{b, s, c}\}$;\\
\ind\ind\ind\ind end if\\
\ind\ind\ind\ind {\it send}($\lambda, \tuple{{\bf p2b}, \vself, \vballotnum}$);\\
\ind\ind\ind end case\\
\ind\ind end switch\\
\ind end for\\
end process
\end{tabular}}
\end{center}\Vex{-2}
  \caption{Acceptor process in vRA Multi-Paxos pseudocode~\cite[Fig.\,4 on
    page 9]{vra15paxos}.}
  \label{fig-vra-acceptor}
\end{figure}
} %end notes

%%%\end{document}

\lncs{} %

\mysec{Mechanically checked proofs in TLAPS}
\label{app-proofs}

\section{TLA+ Specifications and Proofs}
} %end \notes from beginning

%scott: I included here part of the first paragraph above, since annie said we might omit the above section about translation.

\notes{
We manually translated the specification of vRA Multi-Paxos in Figure~\ref{fig-vrapaxos-da} and the two extensions in Section~\ref{sec-optimize} to TLA+. We specified and proved safety for three versions: vRA Multi-Paxos, vRA Multi-Paxos with state reduction, and vRA Multi-Paxos with state reduction and failure detection. The high-level nature of our DistAlgo specifications makes the translation relatively simple.

} %end notes

We developed inductive proofs of safety for all three specifications. %
%Our proofs are manually developed and automatically machine-checked using
%TLAPS~\cite{tlaps15}, a proof system for TLA+.  
%
Like the proof for Multi-Paxos from \cite{Cha+16PaxosTLAPS-FM}, they are
inductive proofs based on several invariants that together imply safety.
The proofs involve three types of invariants: (1) type invariants, stating
that as the system progresses, all data in the system have the expected
types, (2) invariants about local data of processes, for example, about the
values of \texttt{ballot}, \texttt{accepted}, and \texttt{maxb}, and (3)
invariants about global data of the system, in particular, about the
messages sent in the system.

Our proofs for vRA Multi-Paxos differ from the proof for Multi-Paxos from
\cite{Cha+16PaxosTLAPS-FM} for several reasons, including differences
between the algorithms themselves.  For example, the \texttt{accepted} set
in vRA Multi-Paxos contains all triples for which a \co{2a} message was
sent and received and may contain a triple for which a \co{2b} message was not sent, whereas in Multi-Paxos in~\cite{Cha+16PaxosTLAPS-FM}, the \texttt{accepted} set would only keep a triple if a \co{2b} message was sent containing that triple.
%annie: unclear: for a ballot?  this is refering to when the 2b was sent?
%sc: done
Also, to keep our specification in TLA+ close to the specification in
DistAlgo, we model ballots as tuples containing a natural number and a
process ID, not as natural numbers in~\cite{Cha+16PaxosTLAPS-FM}. This
modeling difference has huge impact on the proof, because comparison
operators like $>$ and $\geq$ on natural numbers are built-ins in TLAPS,
and are reasoned about automatically, but comparison operators on tuples need to be defined using predicates, and all of their properties, including fundamental properties like transitivity and non-commutativity, need to be explicitly stated in lemmas and proved.
%annie: quantify "huge".  number of lines added in proof?
In addition, we specify and prove safety of three versions of Multi-Paxos, all of which are variations not considered in~\cite{Cha+16PaxosTLAPS-FM}.

\begin{figure*}[htbp]
  \arxiv{\small}
  \centering
\begin{tabular}{@{~}l@{\ppdp{\hfill}~}|@{~}l@{~}|l@{~\,}l|l@{~\,}l@{~}l@{}}
  & Basic 
  & \multicolumn{2}{@{~}c@{~}}{\mbox{~}\hfill Multi-Paxos\hfill\mbox{~}} 
  & \multicolumn{3}{|l}{vRA Multi-Paxos \& extensions}\\
  \cline{3-7}
%  & \multirow{1}{0pt}{Paxos} & Multi- & Multi- w/ 
%  & \multicolumn{1}{|l}{w/ details} & w/ also & w/ also\\
  Metric & Paxos & Multi- & Multi- w/ 
  & \multicolumn{1}{|l}{w/details} & w/also & w/also\\
  &  & Paxos & Preempt. & \&reconfig. & st. reduct. & fail. detect.\\
  \hline
  Spec size (lines excl.\ comments) & 56 & 81 & 97 & 154 & 157 & 217\\
  Spec size incl.\ comments (lines) & 115 & 133 & 158 & 249 & 254 & 347\\
  \hline
  Proof size (lines excl.\ comments) & 306 & 1003 & 1033 & 4959 & 5005 & 7006\\
  Proof size incl.\ comments (lines) & 423 & 1106 & 1136 & 5256 & 5301 & 7384\\
  \hline
  Max level of proof tree nodes & 7 & 11 & 11 & 12 & 12 & 12\\
  Max degree of proof tree nodes & 3 & 17 & 17 & 28 & 28 & 48\\
  \hline
  \# lemmas & 4 & 11 & 12 & 24 & 23 & 23\\
  \# stability lemmas & 1 & 5 & 6 & 8 & 8 & 8\\
  \# uses of stability lemmas & 8 & 27 & 29 & 76 & 76 & 76\\
  \hline
  \# proofs by induction on set increment & 0 & 4 & 4 & 30 & 30 & 30\\
%  ~~~~set increment\hfill\mbox{} &   &   &   &    &    &   \\
  \# proofs by contradiction & 1 & 1 & 1 & 14 & 16 & 17\\\hline
  \# obligations in TLAPS & 239 & 918 & 959 & 4364 & 4517 & 5580\\\hline
  TLAPS check time (seconds) & 24 & 128 & 94 & 590 & 569 & 781\\
  \hline
\end{tabular}\vspace{-1ex}

\caption{Comparison of results for safety proofs of Basic Paxos
  from~\cite{lam12basicproof}, Multi-Paxos from~\cite{Cha+16PaxosTLAPS-FM},
  and vRA Multi-Paxos.  %Spec size and proof size are measured in lines.
  Spec and proof sizes including comments are also compared because
  they are used in~\cite{Cha+16PaxosTLAPS-FM} as opposed to sizes
  excluding comments.
  Stability lemmas are called continuity lemmas in~\cite{Cha+16PaxosTLAPS-FM}.
  %Width of proof is the product of \textit{Inv} conjuncts and \textit{Next} disjuncts explicitly combined as a goal for the prover to prove; %todo: sum? sc -- product, cartesian product: Inv x Next
  % what is a proof step? sc -- reworded a bit
  % should be $Inv$ and $Next$? - mathmode doesn't compile in this caption - hence, I'm using \textit Haha!!
  % depth is the maximum level of nested steps.
  An obligation is a condition that TLAPS checks.
  The time to check is on an Intel i7-4720HQ 2.6 GHz CPU with 16 GB of
  memory, running Ubuntu 16.04 LTS and TLAPS 1.5.2.}
 \label{fig-sum}
\end{figure*}

\mypar{Results and comparisons}

Figure \ref{fig-sum} presents the results about our specifications and proofs of vRA Multi-Paxos and its extensions, and the specifications and proofs of Multi-Paxos from \cite{Cha+16PaxosTLAPS-FM} and Basic Paxos from~\cite{lam12basicproof}.  First, we compare the specifications and proofs of vRA Multi-Paxos and its extensions with each other:
\begin{itemize}
  \setlength{\itemsep}{1ex}

\item The specification size grows by only 3 lines (1.9\%) from 154 when we
  add state reduction, but by 60 more lines (38\%), for the new actions
  added, when we add failure detection.
  % in lines \texttt{37.1-37.3} and \texttt{38.1-38.2} in
  % Section~\ref{sec-optimize}.

\item The proof size grows by only 46 lines (0.9\%) from 4959 when we add
  state reduction, but by 2001 more lines (40\%) when we add failure
  detection, roughly proportional to the increase in specification size.

\item The maximum level and degree of proof tree nodes remain unchanged
  when state reduction is added.  When failure detection is added, the
  maximum level of proof tree nodes remains unchanged, but the maximum
  degree of proof tree nodes increases by 20 (71\%), from 28 to 48, due to
  more complex proofs for the new actions added for failure detection.
    %annie: but degree goes from 28 to 48 on last column
    %sc- done

\item An interesting decrease of one lemma is seen after state reduction is
  added. The lemma states that the maximum of a set is one of the maximums
  of its two partitions. This lemma was needed in the case when all triples
  in \co{2a} messages are kept by the acceptors. However, owing to state
  reduction, only triples with the maximum ballots are kept, making the
  proofs simpler.

  The number of stability lemmas and their uses remain unchanged when we
  add extensions.  A {\em stability lemma} is a lemma asserting that a
  predicate continues to hold (or not hold) as the system goes from one
  state to the next in a single step.
    
\item The number of proofs by induction on set increment remains unchanged
  when we add extensions. The number of proofs by contradiction
  increases; in those cases, constructive proofs were more challenging.
    
\item The number of obligations, i.e., conditions that TLAPS proves,
  increases by 153 (3.5\%) from 4364 when state reduction is added, and by
  1063 (24\%) more when failure detection is added, contributing to the
  increase in proof size.
    
\item The proof check time decreases by 21 seconds (3.6\%) from 590 to 569
  when state reduction is added. This was expected because, with state
  reduction, for each slot, only the triple with the maximum ballot is
  kept. Upon receiving a triple with a larger ballot,
  %the old one is discarded. Thus, for each slot, at most one triple is kept, 
  only the new triple is kept, 
  and the maximum of a singleton set is the item itself, %. This makes
  making the proof time decrease.
    
  The proof check time increases by 212 seconds (37\%) when failure
  detection is added.  This is expected, because there are more proof
  obligations (24\%) and the proof is larger (40\%).
\end{itemize}

Next, we compare our TLA+ specification and proof of vRA Multi-Paxos
(without state reduction or failure detection) with those of Multi-Paxos
with preemption from \cite{Cha+16PaxosTLAPS-FM}.
\begin{itemize}
  \setlength{\itemsep}{1ex}

\item The specification of vRA Multi-Paxos, excluding comments, is 154
  lines, which is 59\% more.  This increase is
  because~\cite{Cha+16PaxosTLAPS-FM} omits many algorithm details, while
  our specification models the many more details in
  Figure~\ref{fig-vrapaxos-da}. %and reconfiguration.
    
\item The proof of vRA Multi-Paxos, excluding comments, is 4959 lines,
  which is 380\% more. This increase is due to many factors, including more
  actions (for sending \co{2a} messages in two cases and for sending
  decisions), more invariants (about the looser \co{accepted} set and about
  program points for the additional actions), and representing ballots as
  tuples instead of natural numbers, as mentioned above.
    
\item The proof tree for vRA Multi-Paxos is more complex, as shown by the
  65\% increase, from 17 to 28, in the maximum degree of proof tree nodes
  and 9\% increase, from 11 to 12, in the maximum level of proof tree
  nodes.
    
\item Twice as many lemmas are needed for vRA Multi-Paxos, 24 vs.\ 12,
  because properties of operations on tuples need to be explicitly stated
  in lemmas and proved, as mentioned above.
    
  We prove 2 more stability lemmas for vRA Multi-Paxos for the additional
  actions.  The number of uses of stability lemmas increases by 47
  (162\%), from 29 to 76, because of the additional actions and the larger
  number of invariants.  
    
\item The number of proofs by induction on set increment increases by 26
  (650\%), from 4 to 30, the number of proofs by contradiction increases
  from 1 to 14 (1300\%), the number of obligations increases by 3405
  (355\%), from 959 to 4364, and the proof check time increases by 496
  seconds (527\%), from 94 seconds to 590 seconds, all due to increased
  complexity in the specification, more actions, and more invariants.
\end{itemize}

\notes{

}

\newcommand{\ackpeople}{
\arxiv{\subsection*{Acknowledgments}}
\lipics{}
\popl{}
\lncs{}
\icdcs{}
\ppdp{}
We thank Leslie Lamport and Robbert van Renesse for their clear
explanations and helpful discussions about Paxos.  \notanonymous{We thank Bo
  Lin for his robust DistAlgo compiler with excellent support and his
  original DistAlgo program that follows vRA Multi-Paxos pseudocode
  directly.} We thank hundreds of students in distributed algorithms and
distributed systems courses and projects for extending, developing variants
of, testing, evaluating, and model checking our executable specifications,
including running them on distributed machines, in the cloud, etc.  
We thank \notanonymous{Xuetian (Kain) Weng}\anonymous{}
for developing automatic translators from DistAlgo to TLA+.
}
\notanonymous{\ackpeople \ackgrants}

\def\usebib{
\def\bibdir{c:/cygwin64/home/liu/research/bib} %
{
\bibliography{\bibdir/strings,\bibdir/liu,\bibdir/IC,\bibdir/PT,\bibdir/PA,\bibdir/Veri,\bibdir/Lang,\bibdir/Algo,\bibdir/Perform,\bibdir/DB,\bibdir/SE,\bibdir/Sys,\bibdir/Sec,\bibdir/misc,\bibdir/crossref}

\newcommand{\etalchar}[1]{$^{#1}$}
\begin{thebibliography}{GPGMS18}

\bibitem[AS12]{altinbuken2012commodifying}
Deniz Altinbuken and Emin~Gun Sirer.
\newblock Commodifying replicated state machines with openreplica.
\newblock Technical report, Cornell University, 2012.

\bibitem[AvR16]{ovid2016}
Deniz Altinb{\"{u}}ken and Robbert van Renesse.
\newblock Ovid: {A} software-defined distributed systems framework to support
  consistency and change.
\newblock {\em {IEEE} Data Engineering Bulletin}, 39(1):65--80, 2016.

\bibitem[BDFG03]{boichat2003deconstructing}
Romain Boichat, Partha Dutta, Svend Fr{\o}lund, and Rachid Guerraoui.
\newblock Deconstructing paxos.
\newblock {\em ACM SIGACT News}, 34(1):47--67, 2003.

\bibitem[{Ber}12]{BudSandboxPaxos}
{Berkeley Bloom Language Project}.
\newblock Paxos in bud sandbox.
\newblock \url{http://github.com/bloom-lang/bud-sandbox/tree/master/paxos},
  Mar. 13, 2012.

\bibitem[BGG{\etalchar{+}}17]{bobba2017design}
Rakesh Bobba, Jon Grov, Indranil Gupta, Si~Liu, Jos{\'e} Meseguer, Peter~C
  Olveczky, and Stephen Skeirik.
\newblock Design, formal modeling, and validation of cloud storage systems
  using {Maude}.
\newblock Technical report, University of Illinois at Urbana-Champaign, 2017.

\bibitem[BJ87]{birman1987reliable}
Kenneth~P Birman and Thomas~A Joseph.
\newblock Reliable communication in the presence of failures.
\newblock {\em ACM Transactions on Computer Systems (TOCS)}, 5(1):47--76, 1987.

\bibitem[Bur06]{burrows06chubby}
Mike Burrows.
\newblock The {Chubby} lock service for loosely-coupled distributed systems.
\newblock In {\em Proceedings of the 7th USENIX Symposium on Operating Systems
  Design and Implementation}, pages 335--350, 2006.

\bibitem[CGR07]{chandra07paxos}
Tushar~D. Chandra, Robert Griesemer, and Joshua Redstone.
\newblock Paxos made live---{An} engineering perspective.
\newblock In {\em Proceedings of the 26th Annual ACM Symposium on Principles of
  Distributed Computing}, pages 398--407, 2007.

\bibitem[CL18]{ChaLiu18PaxosHistVarTLAPS-NFM}
Saksham Chand and Yanhong~A. Liu.
\newblock Simpler specifications and easier proofs of distributed algorithms
  using history variables.
\newblock In {\em Proceedings of the 10th NASA International Formal Methods
  Symposium, 30 Years of Formal Methods at NASA}, pages 70--86. Springer, 2018.

\bibitem[CLS16]{Cha+16PaxosTLAPS-FM}
Saksham Chand, Yanhong~A. Liu, and Scott~D. Stoller.
\newblock Formal verification of {Multi-Paxos} for distributed consensus.
\newblock In {\em Proceedings of the 21st International Symposium on Formal
  Methods}, pages 119--136. Springer, 2016.

\bibitem[DPLL00]{prisco00revisit}
Roberto De~Prisco, Butler Lampson, and Nancy Lynch.
\newblock Revisiting the {Paxos} algorithm.
\newblock {\em Theoretical Computer Science}, 243(1-2):35--91, 2000.

\bibitem[Erl19]{erlang}
{Erlang Programming Language}.
\newblock \url{http://www.erlang.org/}, 2019.
\newblock Last released May 14, 2019.

\bibitem[FLP85]{fischer85flp}
Michael~J. Fischer, Nancy~A. Lynch, and Michael~S. Paterson.
\newblock Impossibility of distributed consensus with one faulty process.
\newblock {\em Journal of the ACM}, 32(2):374--382, Apr. 1985.

\bibitem[FZWK17]{fonseca2017empirical}
Pedro Fonseca, Kaiyuan Zhang, Xi~Wang, and Arvind Krishnamurthy.
\newblock An empirical study on the correctness of formally verified
  distributed systems.
\newblock In {\em Proceedings of the 12th European Conference on Computer
  Systems}, pages 328--343. ACM Press, 2017.

\bibitem[GLMT09]{georgiou09automated}
Chryssis Georgiou, Nancy~A. Lynch, Panayiotis Mavrommatis, and Joshua~A.
  Tauber.
\newblock Automated implementation of complex distributed algorithms specified
  in the {IOA} language.
\newblock {\em International Journal on Software Tools for Technology
  Transfer}, 11(2):153--171, 2009.

\bibitem[GPGMS18]{garcia18paxos}
{\'A}lvaro Garc{\'\i}a-P{\'e}rez, Alexey Gotsman, Yuri Meshman, and Ilya
  Sergey.
\newblock Paxos consensus, deconstructed and abstracted.
\newblock In {\em European Symposium on Programming}, pages 912--939. Springer,
  2018.

\bibitem[HHK{\etalchar{+}}15]{hawblitzel2015ironfleet}
Chris Hawblitzel, Jon Howell, Manos Kapritsos, Jacob~R. Lorch, Bryan Parno,
  Michael~L. Roberts, Srinath Setty, and Brian Zill.
\newblock {IronFleet}: {P}roving practical distributed systems correct.
\newblock In {\em Proceedings of the 25th Symposium on Operating Systems
  Principles}, pages 1--17. ACM Press, 2015.

\bibitem[Hoa78]{hoare78csp}
C.~A.~R. Hoare.
\newblock Communicating sequential processes.
\newblock {\em Communications of the ACM}, 21(8):666--677, Aug. 1978.

\bibitem[KA08a]{kirsch2008paxosTR}
Jonathan Kirsch and Yair Amir.
\newblock Paxos for system builders.
\newblock Technical Report CNDS-2008-2, John Hopkins University, Mar. 2008.

\bibitem[KA08b]{kirsch2008paxos}
Jonathan Kirsch and Yair Amir.
\newblock Paxos for system builders: {An} overview.
\newblock In {\em Proceedings of the 2nd Workshop on Large-Scale Distributed
  Systems and Middleware}, pages 3:1--3:6. ACM Press, 2008.
\newblock Invited paper.

\bibitem[Kel04]{kellomaki2004ann}
Pertti Kellom{\"a}ki.
\newblock An annotated specification of the consensus protocol of {Paxos} using
  superposition in {PVS}.
\newblock Report~36, Institute of Software Systems, Tampere University of
  Technology, 2004.
\newblock \url{http://www.cs.tut.fi/ohj/laitosraportit/report36-paxos.pdf}.

\bibitem[Lam78]{Lam78}
Leslie Lamport.
\newblock Time, clocks, and the ordering of events in a distributed system.
\newblock {\em Communications of the ACM}, 21(7):558--565, 1978.

\bibitem[Lam96]{lampson1996build}
Butler~W Lampson.
\newblock How to build a highly available system using consensus.
\newblock In {\em International Workshop on Distributed Algorithms}, pages
  1--17. Springer, 1996.

\bibitem[Lam98]{Lam98paxos}
Leslie Lamport.
\newblock The part-time parliament.
\newblock {\em ACM Transactions on Computer Systems}, 16(2):133--169, 1998.

\bibitem[Lam01]{lam01paxos}
Leslie Lamport.
\newblock Paxos made simple.
\newblock {\em SIGACT News (Distributed Computing Column)}, 32(4):51--58, 2001.

\bibitem[Lam02]{Lam02book}
Leslie Lamport.
\newblock {\em Specifying Systems: The TLA+ Language and Tools for Hardware and
  Software Engineers}.
\newblock Addison-Wesley, 2002.

\bibitem[Lam17]{lam-paxos-history}
Leslie Lamport.
\newblock My writings.
\newblock
  \url{http://research.microsoft.com/en-us/um/people/lamport/pubs/pubs.html\#lamport-paxos},
  Accessed Feburary 7, 2017.
\newblock {Lamport's} description of the history of paper~\cite{Lam98paxos}.

\bibitem[Lar09]{larson09erlang}
Jim Larson.
\newblock {Erlang} for concurrent programming.
\newblock {\em Communications of the ACM}, 52(3):48--56, 2009.

\bibitem[LBC16]{lesani2016chapar}
Mohsen Lesani, Christian~J. Bell, and Adam Chlipala.
\newblock Chapar: {Certified} causally consistent distributed key-value stores.
\newblock In {\em Proceedings of the 43rd Annual ACM SIGPLAN-SIGACT Symposium
  on Principles of Programming Languages}, pages 357--370. ACM Press, 2016.

\bibitem[LL18]{distalgo18git}
Bo~Lin and Yanhong~A. Liu.
\newblock {DistAlgo}: A language for distributed algorithms.
\newblock \url{http://github.com/DistAlgo}, 2018.
\newblock Beta release September 27, 2014. Latest release September 18, 2018.

\bibitem[LLS17]{distalgo17lang}
Yanhong~A. Liu, Bo~Lin, and Scott Stoller.
\newblock {{DistAlgo} Language Description}.
\newblock \url{http://distalgo.cs.stonybrook.edu}, Mar. 2017.

\bibitem[LMD14]{lam12basicproof}
Leslie Lamport, Stephan Merz, and Damien Doligez.
\newblock {A TLA spefication of the Paxos Consensus algorithm described in
  Paxos Made Simple and a TLAPS-checked proof of its correctness}.
\newblock file {\tt /tlapm/examples/paxos/Paxos.tla} in TLAPS distribution
  \url{http://tla.msr-inria.inria.fr/tlaps/dist/current/tlaps-1.4.3.tar.gz},
  November 2012. Last modified November 28, 2014.

\bibitem[LSCW18]{Liu+8DistInv-TLA}
Yanhong~A. Liu, Scott~D. Stoller, Saksham Chand, and Xuetian Weng.
\newblock Invariants in distributed algorithms.
\newblock In {\em Proceedings of the TLA+ Community Meeting}, Oxford, U.K.,
  2018.

\bibitem[LSL12]{Liu+12DistSpec-SSS}
Yanhong~A. Liu, Scott~D. Stoller, and Bo~Lin.
\newblock High-level executable specifications of distributed algorithms.
\newblock In {\em Proceedings of the 14th International Symposium on
  Stabilization, Safety, and Security of Distributed Systems}, pages 95--110.
  Springer, 2012.

\bibitem[LSL17]{Liu+17DistPL-TOPLAS}
Yanhong~A. Liu, Scott~D. Stoller, and Bo~Lin.
\newblock From clarity to efficiency for distributed algorithms.
\newblock {\em ACM Transactions on Programming Languages and Systems},
  39(3):12:1--12:41, May 2017.

\bibitem[LSLG12]{Liu+12DistPL-OOPSLA}
Yanhong~A. Liu, Scott~D. Stoller, Bo~Lin, and Michael Gorbovitski.
\newblock From clarity to efficiency for distributed algorithms.
\newblock In {\em Proceedings of the 27th ACM SIGPLAN Conference on
  Object-Oriented Programming, Systems, Languages and Applications}, pages
  395--410. ACM Press, 2012.

\bibitem[{Mic}15]{tlaps15}
{Microsoft Research-Inria Joint Center}.
\newblock {TLA+ Proof System (TLAPS)}.
\newblock \url{http://tla.msr-inria.inria.fr/tlaps/}, Last released June 2015.

\bibitem[{Mic}18]{tlatoolbox18}
{Microsoft Research}.
\newblock {The TLA Toolbox}.
\newblock \url{http://lamport.azurewebsites.net/tla/toolbox.html}, Last
  modified January 30, 2018.

\bibitem[Mil80]{milner80ccs}
Robin Milner.
\newblock {\em A Calculus of Communicating Systems}, volume~92.
\newblock Springer, 1980.

\bibitem[OL88]{oki88vsr}
Brian~M Oki and Barbara~H Liskov.
\newblock Viewstamped replication: {A} new primary copy method to support
  highly-available distributed systems.
\newblock In {\em Proceedings of the 7th Annual ACM Symposium on Principles of
  Distributed Computing}, pages 8--17. ACM Press, 1988.

\bibitem[OO14]{ongaro14raft}
Diego Ongaro and John Ousterhout.
\newblock In search of an understandable consensus algorithm.
\newblock In {\em 2014 USENIX Annual Technical Conference}, pages 305--319.
  USENIX Association, 2014.

\bibitem[PHL{\etalchar{+}}17]{padon2017reducing}
Oded Padon, Jochen Hoenicke, Giuliano Losa, Andreas Podelski, Mooly Sagiv, and
  Sharon Shoham.
\newblock Reducing liveness to safety in first-order logic.
\newblock {\em Proceedings of the ACM on Programming Languages}, 2(POPL):26,
  2017.

\bibitem[PLSS17]{padon2017paxos}
Oded Padon, Giuliano Losa, Mooly Sagiv, and Sharon Shoham.
\newblock Paxos made {EPR}: {Decidable} reasoning about distributed protocols.
\newblock {\em Proceedings of the ACM on Programming Languages}, 1(OOPSLA):108,
  2017.

\bibitem[TLM{\etalchar{+}}18]{taube2018modular}
Marcelo Taube, Giuliano Losa, Kenneth~L. McMillan, Oded Padon, Mooly Sagiv,
  Sharon Shoam, James R.Wilcox, and Doug Woos.
\newblock Modularity for decidability of deductive verification with
  applications to distributed systems.
\newblock In {\em Proceedings of the 2018 ACM SIGPLAN Conference on Programming
  Language Design and Implementation}, pages 662--677, 2018.

\bibitem[vRA15]{vra15paxos}
Robbert van Renesse and Deniz Altinbuken.
\newblock Paxos made moderately complex.
\newblock {\em ACM Computing Surveys}, 47(3):42:1--42:36, Feb. 2015.

\bibitem[vRSS15]{van15vive}
Robbert van Renesse, Nicolas Schiper, and Fred~B. Schneider.
\newblock Vive la diff{\'e}rence: {Paxos} vs.\ viewstamped replication vs.\
  {Zab}.
\newblock {\em IEEE Transactions on Dependable and Secure Computing},
  12(4):472--484, July 2015.

\bibitem[WWA{\etalchar{+}}16]{woos2016planning}
Doug Woos, James~R Wilcox, Steve Anton, Zachary Tatlock, Michael~D Ernst, and
  Thomas Anderson.
\newblock Planning for change in a formal verification of the {Raft} consensus
  protocol.
\newblock In {\em Proceedings of the 5th ACM SIGPLAN Conference on Certified
  Programs and Proofs}, pages 154--165, 2016.

\bibitem[WWP{\etalchar{+}}15]{wilcox2015verdi}
James~R. Wilcox, Doug Woos, Pavel Panchekha, Zachary Tatlock, Xi~Wang,
  Michael~D. Ernst, and Thomas Anderson.
\newblock Verdi: A framework for implementing and formally verifying
  distributed systems.
\newblock In {\em Proceedings of the 36th ACM SIGPLAN Conference on Programming
  Language Design and Implementation}, pages 357--368. ACM Press, 2015.

\end{thebibliography}
\arxiv{\bibliographystyle{alpha}}%
\lipics{}%
\lncs{}
\icdcs{}
}
}
\usebib

\icdcs{}
\end{document}